\documentclass[11point]{article}
\usepackage{amssymb}
\usepackage{amsmath}
\usepackage{paralist}
\usepackage{graphics}
\usepackage{epsfig}
\usepackage{graphicx}
\usepackage{epstopdf}
\usepackage[colorlinks=true]{hyperref}
\hypersetup{urlcolor=blue, citecolor=red}
\usepackage{algorithm}
\usepackage{algorithmic}

\newtheorem{example}{Example}

\title{Edge-adaptive $\ell_2$ regularization image reconstruction from non-uniform Fourier data}

\begin{document}
\maketitle

\centerline{\scshape Victor Churchill$^*$}
\medskip
{\footnotesize
 \centerline{Department of Mathematics}
 \centerline{Dartmouth College}
 \centerline{Hanover, NH 03755, USA}
 \centerline{victor.a.churchill.gr@dartmouth.edu}
}

\medskip

\centerline{\scshape Rick Archibald}
{\footnotesize
 \centerline{Computer Science and Mathematics Division}
 \centerline{Oak Ridge National Laboratory}
 \centerline{Oak Ridge, TN 37830, USA}
 \centerline{archibaldrk@ornl.gov}
}

\medskip

\centerline{\scshape Anne Gelb}
{\footnotesize
 \centerline{Department of Mathematics}
   \centerline{Dartmouth College}
   \centerline{Hanover, NH 03755, USA}
   \centerline{annegelb@math.dartmouth.edu}
\bigskip

\begin{abstract}
Total variation regularization based on the $\ell_1$ norm is ubiquitous in image reconstruction. However, the resulting reconstructions are not always as sparse in the edge domain as desired. Iteratively reweighted methods provide some improvement in accuracy, but at the cost of extended runtime. In this paper we examine these methods for the case of data acquired as non-uniform Fourier samples. We then develop a non-iterative weighted regularization method that uses a pre-processing edge detection to find exactly where the sparsity should be in the edge domain. We show that its performance in terms of both accuracy and speed has the potential to outperform reweighted TV regularization methods.
\end{abstract}

\section{Introduction}

Data for reconstruction of piecewise smooth functions and images are sometimes acquired as non-uniform Fourier samples. This is the case in magnetic resonance imaging (MRI) and synthetic aperture radar (SAR). Since sparsity is inherent in the edge domain of piecewise smooth functions and images, $\ell_1$ based total variation (TV) regularization, \cite{rudin1992nonlinear}, is commonly employed for reconstruction. The development of compressed sensing, \cite{candes2006robust,candes2006stable,candes2006near,donoho2006compressed}, has provided theoretical justification for using $\ell_1$ regularization to promote sparsity in the appropiate domain. In some instances, however, these reconstructions are not as sparse in the edge domain as desired. This may be due to non-uniform sampling, noise, or the fact that the TV transform is not actually sparsifying with respect to a particular function, and is likely a combination of these factors. Consequently, the  overall accuracy is reduced. One popular approach for correcting this problem is to use an iterative reweighting scheme, \cite{candes2008enhancing,chartrand2008iteratively,daubechies2010iteratively,gorodnitsky1997sparse,mansour2012support,wang2010sparse,wipf2010iterative}, which employs multiple runs of weighted $\ell_p$ minimization (typically $p = 1$ but $p = 2$ is also an option).  The main idea is to find the locations of non-zero entries in the edge domain and then apply regularization away from those locations. Iterative reweighting methods are shown to be more accurate than single-run TV methods, \cite{candes2008enhancing,chartrand2008iteratively}, and have been applied to problems where data are acquired as uniform Fourier samples, \cite{candes2008enhancing}.  The extension of these algorithms is straightforward for non-uniform Fourier data acquisition, although the implementation requires a non-uniform fast Fourier transform (NFFT), \cite{dutt1993fast,greengard2004accelerating,lee2005type}, and there are additional errors corresponding to the resulting fidelity term.\footnote{This is the case whenever the acquired data are non-uniform Fourier samples, see e.g. \cite{dutt1993fast}.}  It is important to note that in this investigation we are considering that the data acquired are noisy {\em continuous} Fourier samples, which means that using the discrete NFFT generates additional model mismatch, \cite{archibald2016image}.

This paper provides an alternative approach to this problem. We propose an algorithm for image reconstruction from non-uniform Fourier data that, rather than using iterative weighting, uses edge detection to indicate regions of sparsity in the edge domain and targets weighted $\ell_2$ regularization appropriately. Unlike iterative reweighting, which requires multiple iterations of $\ell_1$ minimization, there are only two steps to our method. The first step uses an $\ell_1$ regularization based edge detection to create a mask, i.e.~a weighting matrix, which dictates where non-zero entries are expected in the edge domain. The second step uses this mask to target $\ell_2$ regularization only in non-edge regions, that is, regions of the function or image that are truly sparse in the edge domain. Put another way, our method uses regularization on targeted areas that we actually expect to be zero in the edge domain. Therefore, with a properly chosen mask, it is appropriate to regularize using the $\ell_2$ norm, making the algorithm much more cost efficient. Moreover, in regions containing edges, our method relies solely on the fidelity term. This approach is particularly advantageous when noise is added since we can weigh the fidelity term lightly against the regularization term, which encourages noise reduction in non-edge regions. We call this method edge-adaptive $\ell_2$ regularization.

There are several benefits to our proposed algorithm. First, it compares favorably in terms of accuracy (pointwise error) to iteratively weighted $\ell_1$ regularization methods. It also provides better resolution around jumps than these methods. It is also more efficient to implement.  Reweighted $\ell_1$ methods take multiple iterations to identify the  sparse regions. Our method uses only a single $\ell_1$ minimization in the pre-processing edge detection step followed by a single $\ell_2$ minimization in the main reconstruction step. Further, we are able to use faster conjugate gradient descent optimization methods available for $\ell_2$ regularized problems. Finally, there is a closed form solution to our problem, which may be valuable in some settings.

The rest of the paper is organized as follows: Section \ref{Preliminaries} covers the necessary background in image reconstruction from non-uniform Fourier data. Section \ref{IR} applies an iteratively reweighted $\ell_1$ regularization method to this problem. Section \ref{EAR} describes the edge-adaptive approach and its benefits over the iterative method. Section \ref{numerics} looks at numerical results. Conclusions and future work are in Section \ref{conclusion}.

\section{Preliminaries} \label{Preliminaries}
In the one-dimensional case, we consider a piecewise smooth function $f:[-1,1]\rightarrow\mathbb{R}$. Suppose we are given a finite sequence of non-uniform Fourier samples of $f$,
\begin{eqnarray}
\label{eq:fourierdata}
\hat{f}(\lambda_k) & = & \frac{1}{2}\int_{-1}^1f(x)e^{-\pi i\lambda_kx}dx,
\end{eqnarray}
where $\lambda_k\in\mathbb{R}$ and $k=-M,\ldots,M$. Specifically, we look at jittered sampling, defined by
\begin{eqnarray}
\label{eq:jittered}
\lambda_k & = & k-\left\lfloor \frac{2M+1}{2} \right\rfloor -1+\frac{1-2\xi_k}{4},
\end{eqnarray}
where $\xi_k \sim U([0,1])$. We will also consider the case where the underlying Fourier data in (\ref{eq:fourierdata}) are noisy, given by
\begin{equation}
\label{eq:noisydata}
\hat{f}^\eta (\lambda_k) = \hat{f}(\lambda_k) + \eta_k,
\end{equation}
for $k = -M,\cdots,M$.  Here $\eta_k\sim\mathcal{CN}(0,\sigma^2)$, meaning $\eta_k$ is a complex Gaussian random variable with mean $0$ and variance $\sigma^2$.

In two dimensions, we analogously consider the piecewise smooth function $f:[-1,1]^2\rightarrow \mathbb{R}$. Suppose we are given a finite sequence of non-uniform Fourier samples of $f$,
\begin{eqnarray}
\label{eq:fourierdata2d}
\hat{f}(\mathbf{\lambda}_{\mathbf{k}}) & = & \frac{1}{4}\int_{-1}^1\int_{-1}^1 f(x,y)e^{-\pi i\lambda_{k_1}x}e^{-\pi i\lambda_{k_2}y}dxdy,
\end{eqnarray}
where $\{\mathbf{\lambda}_{\mathbf{k}} = (\lambda_{k_1},\lambda_{k_2}):k_1,k_2=-M,\ldots,M\}\in\mathbb{R}^2$. The non-uniform jittered sampling pattern for $\lambda_{\mathbf{k}}$ is given by
\begin{eqnarray}
\label{eq:jittered2d}
\lambda_{\mathbf{k}} & = &\mathbf{k}-\left\lfloor \frac{2M+1}{2} \right\rfloor -1+\frac{1-2\mathbf{\xi}_{\mathbf{k}}}{4},
\end{eqnarray}
where $\mathbf{\xi}_{\mathbf{k}} \sim U([0,1])^2$. The sampling patterns in (\ref{eq:jittered}) and (\ref{eq:jittered2d}), displayed in Figure \ref{fig:nonuniformgrid}, simulate Cartesian grid samples with slight deviations that sometimes occur in real world measurement systems. We will also consider noisy two-dimensional Fourier data, $\hat{f}^\eta(\lambda_{\mathbf{k}})$, defined analogously to (\ref{eq:noisydata}).

\begin{figure}[h!]
\centering
\includegraphics[width=.3\textwidth]{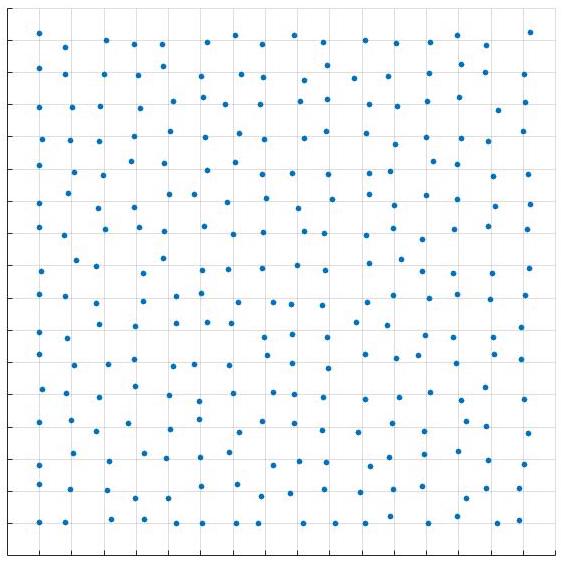} 
\caption{Non-uniform sampling $\mathbf{\lambda}_{\mathbf{k}}$ as in (\ref{eq:jittered2d}).}
\label{fig:nonuniformgrid}
\end{figure}

For ease of presentation, we begin by describing some known techniques for piecewise smooth function reconstruction in the one-dimensional case, where the acquired data are given in (\ref{eq:fourierdata}). These methods are easily extended to reconstruct two-dimensional images, which will be demonstrated in Section \ref{IR}.

Let $\mathbf{f}=\{f(x_j):j=-J,\ldots,J\}$ and $\mathbf{\hat{f}}=\{\hat{f}(\lambda_k):k=-M,\ldots,M\}$. Since the underying function $f$ is piecewise smooth, it is sparse in the edge domain. Hence $\ell_1$ regularization provides an effective means for its reconstruction. In particular,  $f$  can be determined on a set of discrete grid points by solving the unconstrained optimization problem given by
\begin{align}
\label{eq:l1reg}
\mathbf{f}^* &= \arg \min_{\mathbf{g}} \left(||\mathcal{F}_N\mathbf{g} - \mathbf{\hat{f}}||_2^2 + \lambda ||\mathbf{T}\mathbf{g}||_1\right).
\end{align}
Here $\mathcal{F}_{N}$ is the NFFT matrix (see e.g.~\cite{greengard2004accelerating,lee2005type} for details about NFFT solvers), $\lambda>0$ is the $\ell_1$ regularization parameter, and $\mathbf{T}$ is a transformation to the edge domain. The choice for regularization parameter, $\lambda$,  is typically problem dependent, \cite{osher2005iterative}.  We note that in this investigation we used single digit accuracy for the NFFT algorithm.

If $f$ is a piecewise constant, for example a cross section of the Shepp Logan phantom (with increased contrast for visual perception) seen in Figure \ref{fig:shepplogan}, then a standard choice for reconstruction is the solution to the TV-regularized optimization problem
\begin{align}
\label{eq:TV}
\mathbf{f}^* &=\arg\min_{\bf g} \left(||{\mathcal F}_{N}{\bf g} - {\bf \hat f}||_2 +\lambda \sum_{j=-J}^{J-1}\left|\mathbf{g}_{j+1}-\mathbf{g}_j\right|\right),
\end{align}
which is frequently written as
\begin{align}
\label{eq:TVmat}
\mathbf{f}^* &=\arg\min_{\bf g} \left(||{\mathcal F}_{N}{\bf g} - {\bf \hat f}||_2 +\lambda ||\mathbf{D}\mathbf{g}||_1\right).
\end{align}
Here $\mathbf{D}$ is simply the  matrix that encodes the entry information from the sum in (\ref{eq:TV}). Note that  the sparsifying transformation in (\ref{eq:TVmat}) is an approximation to the first derivative, effectively penalizing high gradients in the function and therefore encouraging sparsity in the edge domain. Applying TV regularization causes the well known staircase effect, whereby the solution is held to be piecewise constant regardless of the smoothness of the underlying function. If $f$ is a sparse signal, e.g.~a spike train, then a standard reconstruction is the solution to the $\ell_1$ regularized optimization problem
\begin{align}
\label{eq:sparse}
\mathbf{f}^* &=\arg\min_{\bf g} \left(||{\mathcal F}_{N}{\bf g} - {\bf \hat f} ||_2 +\lambda ||{\bf g}||_1\right).
\end{align}
If $f$ can be modeled as a piecewise polynomial, then a suitable regularization choice is high order total variation (HOTV), \cite{archibald2016image,chan2000high}, yielding
\begin{align}
\label{eq:HOTV}
\mathbf{f}^* &=\arg\min_{\bf g} \left(||{\mathcal F}_{N}{\bf g} - {\bf \hat f}||_2 +\lambda ||L^m{\bf g}||_1\right).
\end{align}
Here $L^m$ is the $m^{th}$ order polynomial annihilation (PA) transform, \cite{archibald2016image,archibald2005polynomial}.\footnote{Although there are subtle differences in the derivations and normalizations, the PA transform can be thought of as a variant  of  HOTV. Because part of our investigation discusses parameter selection, which depends explicitly on $||L^m f||$, we will exclusively use the PA transform as it appears in \cite{archibald2016image} so as to avoid any confusion. Explicit formulations for the PA transform matrix can also be found in \cite{archibald2016image}.  We further note that the method we develop here can be easily adapted for other sparsifying transformations.}   For example, when $m = 3$ we have 
\begin{eqnarray}
\label{eq:Lm}
 L^3 & = & \begin{bmatrix}
-\frac12 & \frac32 & -\frac32 & \frac12 & 0 & \cdots & 0\\
0 &-\frac12 & \frac32 & -\frac32 & \frac12 & \cdots & 0\\
\vdots & & \ddots & \ddots & \ddots & \ddots & \\
0&\cdots &0 & -\frac12 &\frac32&-\frac32&\frac12
 \end{bmatrix}.
\end{eqnarray}
In general, $L^m$ can be viewed as a normalized appproximation of the $m^{th}$ derivative. In particular, $L^0=I$, i.e.~(\ref{eq:HOTV}) is equivalent to (\ref{eq:sparse}), while using $L^1$ yields (\ref{eq:TV}).

For two-dimensional images, we regularize in the $x$ and $y$ directions separately and solve
\begin{align}
\label{eq:HOTV2d}
\mathbf{f}^* &=\arg\min_{\bf g} \left(||{\mathcal F}_{N}{\bf g} - {\bf \hat f}||_2 +\lambda \left(||L^m{\bf g}||_1+||\mathbf{g}(L^m)^T||_1\right)\right),
\end{align}
where the regularization term $||L^m\mathbf{g}||_1$ penalizes gradients in the $x$ direction and the regularization term $||\mathbf{g}(L^m)^T||_1$ penalizes gradients in the $y$ direction.\footnote{The PA matrix $L^m$ can be constructed for two dimensional images, \cite{archibald2005polynomial}.  However, in \cite{archibald2016image} it was demonstrated that splitting the dimensions was more cost effective and did not reduce the quality of the reconstruction.}

As noted previously, using (\ref{eq:HOTV}) is effective in reconstructing piecewise smooth functions and images in a large number of applications.  For a variety of reasons, however, the assumption that the function or image is sparse in the edge domain is often flawed. One reason is noise, which will immediately degrade the edge sparsity of the solution to (\ref{eq:HOTV}). For TV regularization, the assumption that the transformed image is sparse is often inadequate due to smooth variation away from jumps. This is somewhat mitigated by HOTV regularization. However, if due to lack of resolution the image has variation not accounted for away from discontinuities, even high order transformations will not produce the desired sparsity. Another source of error is non-uniform sampling, since some compromise in accuracy is necessary to maintain the efficiency of the NFFT. Finally, all $\ell_1$ based methods suffer from the fact that the $\ell_1$ norm penalizes large magnitudes more heavily.

A popular approach to mitigating error from issues such as noise, lack of resolution, and magnitude dependence is to use a scheme that employs iteratively reweighted (IR) regularization, \cite{candes2008enhancing,chartrand2008iteratively,daubechies2010iteratively,gorodnitsky1997sparse,mansour2012support,wang2010sparse,wipf2010iterative}. In these methods, multiple passes of TV or HOTV regularization with weighted $\ell_p$ norms are used to ``narrow in'' on spikes in the edge domain. The weight at each point on the spatial grid is typically inversely proportional to the magnitude of that point in the edge domain of the previous iteration. That is, the regularization is more strongly enforced at points deemed as non-jumps and more weakly enforced at those identified as jumps. In this way, iterative reweighting more democratically penalizes high gradients and regularizes based on the spatial distribution of the sparsity. These iterative methods are typically more accurate than single pass methods, \cite{candes2008enhancing,chartrand2008iteratively}. We will use the next section to discuss IR methods in more detail.

\section{Iteratively reweighted regularization methods}\label{IR}

As explained in \cite{candes2006robust}, reconstructing an image or function via solving
\begin{align}
\label{eq:l0}
\mathbf{f}^*&=\arg\min_{\mathbf{g}} \left(||\mathcal{F}_N\mathbf{g}-\mathbf{\hat{f}}||_2^2+\lambda||L^m\mathbf{g}||_0\right),
\end{align}
where $||\cdot||_0$ counts non-zero values, promotes the most sparsity in the edge domain of the image. However, this combinatorial problem is NP-hard. As in (\ref{eq:HOTV}), the $\ell_1$ term acts as a convex surrogate for the $\ell_0$ term, making the problem easier to solve. But it does not encourage sparsity in the edge domain as much. Naturally, this begs the question of whether there are better surrogates that generate solvable optimization problems.

The approach of \cite{candes2008enhancing} is to regularize using the log-sum function, a concave penalty function that more closely resembles the $\ell_0$ norm and is therefore more sparsity-inducing. That reconstruction would be the solution to the optimization problem
\begin{align}
\label{eq:logsum}
\mathbf{f}^* &=\arg\min_\mathbf{g} \left(||\mathcal{F}_N\mathbf{g}-\mathbf{\hat{f}}||_2^2+\sum_{j=-J}^J \log(|(L^m\mathbf{g})_{j}|+\epsilon)\right),
\end{align}
where $\epsilon>0$ is a parameter to stay within the domain of the logarithm. Since the log-sum function is nonconvex, (\ref{eq:logsum}) is difficult solve. Instead, we can approximate it with a series of weighted $\ell_1$ based minimizations of the form
\begin{align}
\label{eq:l0_log}
\mathbf{f}^*&=\arg\min_{\mathbf{g}} \left(||\mathcal{F}_N\mathbf{g}-\mathbf{\hat{f}}||_2^2+\lambda||WL^m\mathbf{g}||_1\right),
\end{align}
where $W$ is a diagonal matrix of weights. The main idea is that large weights can be used to discourage non-zero entries in the edge domain, while small weights can be used to encourage non-zero entries. Hence this method will penalize non-zero edge domain magnitudes more fairly, removing the magnitude dependence of unweighted $\ell_1$ regularization. To achieve this, weights inversely proportional to the edge domain magnitudes of the previous iteration are used. In this way, IR methods encourage sparsity in the edge domain by regularizing less in areas with jumps and more in areas without jumps.

Algorithms \ref{alg:irl1} and \ref{alg:irl12d} are modified versions of the iterative weighting method used in \cite{candes2008enhancing} for one- and two-dimensional functions, respectively. They include the PA transform $L^{m}$  so that the staircasing effect from standard TV can be avoided. When $m=1$, we will refer to Algorithms \ref{alg:irl1} and \ref{alg:irl12d} as reweighted total variation (RWTV) methods. For $m\ge2$, they will be called reweighted high order total variation (RWHOTV) methods.

\begin{algorithm}[htbp!]
\caption{Iteratively reweighted (IR) $\ell_1$ regularization reconstruction in one dimension}
\label{alg:irl1}
\begin{algorithmic}[1]
\STATE Set $\ell=0$ and $w_{j}^{(0)}=1$ for $j=-J,\ldots,J$.  Fix the regularization parameter $\rho>0$, the weighting parameter $\epsilon>0$, and an appropriate total variation order $m$.

\STATE Solve the weighted regularization minimization problem
\begin{align}\label{eq:irl1}
\mathbf{f}^{(\ell)} &=\arg\min_{\mathbf{g}}\left(||\mathcal{F}_{N}\mathbf{g}-\mathbf{\hat{f}}||_2^2+\rho||W^{(\ell)}L^m\mathbf{g}||_1\right)
\end{align}
where $W^{(\ell)}= \text{diag}(w^{(\ell)})$.

\STATE Update the weights. For each $j=-J,\ldots,J$,
\begin{align}\label{eq:weights}
w_{j}^{(\ell+1)} &= \frac{1}{|(L^m\mathbf{f}^{(\ell)})_{j}|+\epsilon}.
\end{align}

\STATE Terminate on convergence or when $\ell$ attains a pre-specified maximum number of iterations $\ell_{max}$. Otherwise, increment $\ell$ and go to step $2$.

\end{algorithmic}
\end{algorithm}

\begin{algorithm}[htbp!]
\caption{Iteratively reweighted (IR) $\ell_1$ regularization reconstruction in two dimensions}
\label{alg:irl12d}
\begin{algorithmic}[1]
\STATE Set $\ell=0$ and $v_{i,j}^{(0)}=1$ for $i,j=-J\ldots J$ and $w_{i,j}^{(0)}=1$ for $i,j=-J\ldots J$.   Fix the regularization parameter $\rho>0$, the weighting parameter $\epsilon>0$, and an appropriate total variation order $m$.

\STATE Solve the weighted regularization minimization problem
\begin{equation}\label{eq:irl12d}
\begin{split}
\mathbf{f}^{(\ell)} &=\arg\min_{\mathbf{g}}\bigg\{||\mathcal{F}_{N}\mathbf{g}-\mathbf{\hat{f}}||_2^2\\
&+\rho\left(\sum_{i=-J}^J\sum_{j=-J}^J v_{i,j}^{(\ell)}|(L^m\mathbf{g})_{i,j}|+\sum_{i=-J}^J\sum_{j=-J}^J w_{i,j}^{(\ell)}|(\mathbf{g}(L^m)^T)_{i,j}|\right)\bigg\}
\end{split}
\end{equation}

\STATE Update the weights. For each $(i,j)$ such that $i,j=-J,\ldots,J$,
\begin{align}\label{eq:weights2d}
v_{i,j}^{(\ell+1)} &= \frac{1}{|(L^m\mathbf{f}^{(\ell)})_{i,j}|+\epsilon}\quad\text{and}\quad w_{i,j}^{(\ell+1)}= \frac{1}{|(\mathbf{f}^{(\ell)}(L^m)^T)_{i,j}|+\epsilon}.
\end{align}

\STATE Terminate on convergence or when $\ell$ attains a pre-specified maximum number of iterations $\ell_{max}$. Otherwise, increment $\ell$ and go to step $2$.

\end{algorithmic}
\end{algorithm}

As explained in \cite{candes2008enhancing}, the main advantages of this method are increased accuracy using the same number of Fourier coefficients and removal of magnitude dependence of the unweighted $\ell_1$ norm. A chief example of when this method works very well is found in Section $3.6$, in particular Figure $10$, of \cite{candes2008enhancing}. These advantages are balanced with some disadvantages. The runtime is increased $\ell_{max}$ times for this method, since each iteration requires an $\ell_1$ minimization step. In addition, this method introduces another parameter $\epsilon$. No comprehensive method for choosing this parameter is provided in \cite{candes2008enhancing} and the success of this algorithm depends on an appropriate choice. Finally, there still appear to be clear sources of error generated by this method.

As prototype examples to test the IR method, we consider
\begin{example}
\label{ex:f1}
\begin{align}
\label{eq:f1}
f_1(x) &= \left\{\begin{array}{cc}
\cos(x/2) & x \ge0\\
-\cos(x/2) & x<0
\end{array}\right.,
\end{align}
\end{example}
\begin{example}
\label{ex:f2}
\begin{align}
\label{eq:f2}
f_2(x) &= \left\{\begin{array}{cc}
\frac32 & -\frac{3\pi}{4}\le x<-\frac{\pi}{2} \\
\frac74-\frac x2+\sin(7x-\frac14) & -\frac\pi4\le x<\frac\pi8 \\
\frac{11x}{4}-5 & \frac{3\pi}{8}\le x<\frac{3\pi}{4} \\
0 & \text{else}
\end{array}\right.,
\end{align}
\end{example}
and
\begin{example}
\label{ex:example3}
\begin{align}
\label{eq:f2D}
f_3(x,y) &= \left\{\begin{array}{cc}
\cos(\pi (x^2+y^2)) & x^2+y^2\le\frac12\\
\cos(\pi (x^2+y^2) - \frac\pi2) & x^2+y^2>\frac12
\end{array}\right..
\end{align}
\end{example}

One issue with Algorithms \ref{alg:irl1} and \ref{alg:irl12d} is that when noise is present there are non-zero weights being applied in areas of smooth variation (and no variation), causing false jump identifications and ultimately oscillation in the reconstruction. Effectively, the oscillations caused by noise in the initial TV solution are propagated through all the iterations via the weighting matrix. The algorithm has no way to validate whether these oscillations are from actual variation, a jump, or noise. Figure \ref{fig:irl1noise} demonstrates the use of Algorithm \ref{alg:irl1} for $f_1(x)$ and $f_2(x)$ when the given Fourier data (\ref{eq:fourierdata}) is noise free and when complex zero-mean Gaussian noise is added. Notice how the oscillations in the reconstruction increase where the function has more smooth variation.

\begin{figure}[htbp!]
\begin{center}
\includegraphics[width=.6\textwidth]{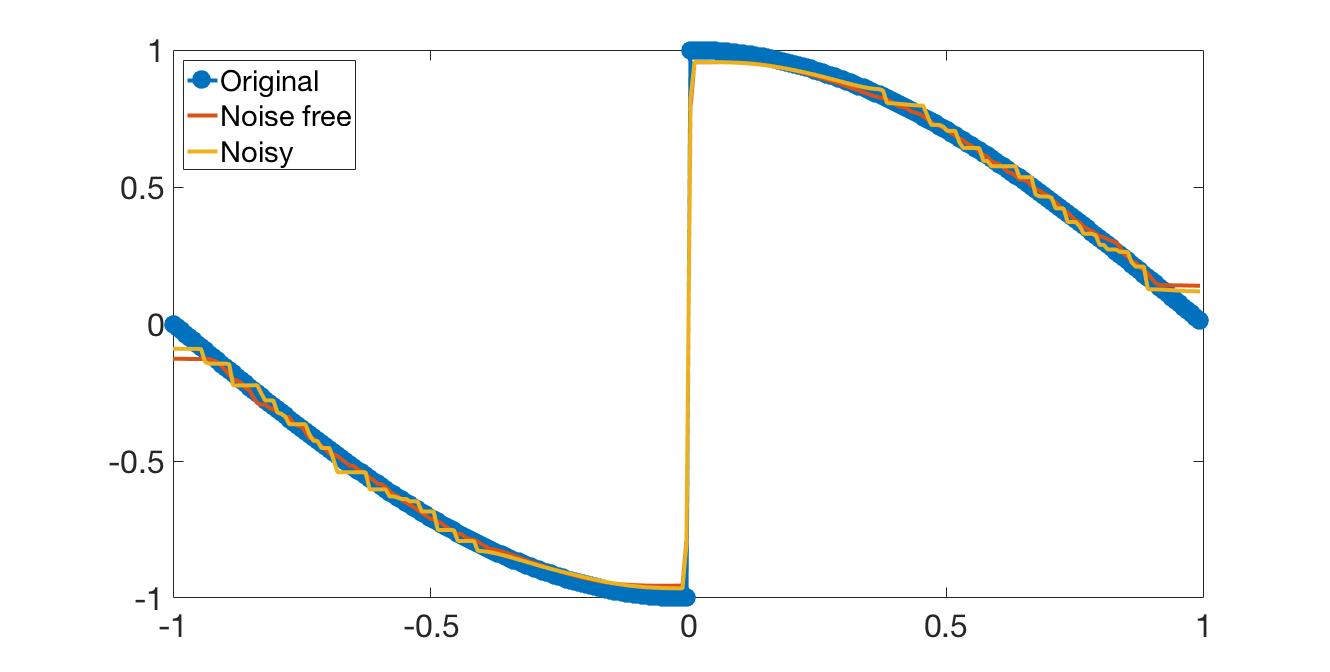} 
\includegraphics[width=.6\textwidth]{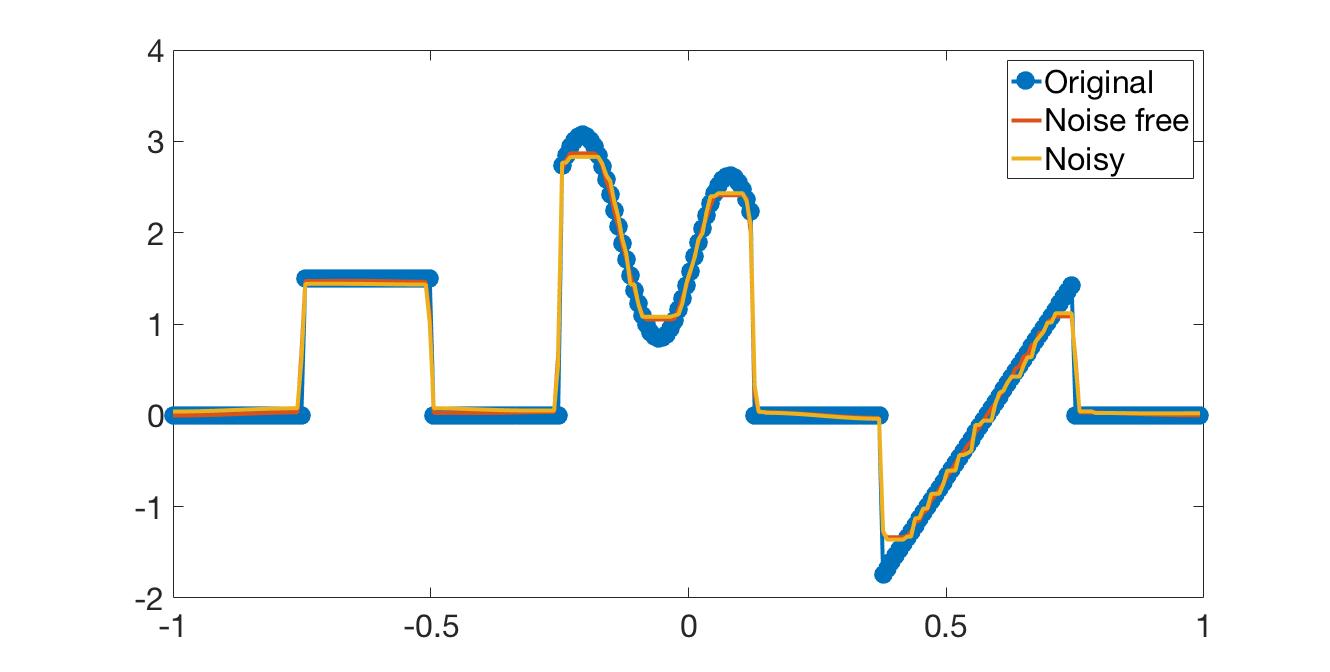} 
\end{center}
\caption{$f_1(x)$ (top) and $f_2(x)$ (bottom) reconstructed via Algorithm \ref{alg:irl1} using $m=1$ (standard TV) from $257$ Fourier modes on $257$ grid points. The red reconstructions are noise-free as in (\ref{eq:fourierdata}), while the yellow have zero-mean complex Gaussian noise added to the Fourier coefficients as in (\ref{eq:noisydata}). Here we use a signal to noise ratio (SNR) of $20$ dB. For $f_1(x)$, we used parameters $\rho=1$, $\ell_{max}=25$, and $\epsilon=1.9$. For $f_2(x)$, we used parameters $\rho=1$, $\ell_{max}=25$, and $\epsilon=2.9$.}
\label{fig:irl1noise}
\end{figure}

The source of this error is from points weighted between $0$ and $\frac{1}{\epsilon}$. These weights can indicate either a small jump or variation in a smooth region that is beyond the resolution of the problem, which could be attributable to noise, or more simply the variation of the function itself. If there is a small jump, the iterative reweighting strategy still regularizes at that point, albeit relatively less. If it is just noise or normal variation, then the algorithm regularizes less at that point for no reason. This will automatically reduce the algorithm's ability to separate the true scales of the underlying image by causing false jump identifications, leading to an overall less accurate reconstruction.

\begin{figure}[h!]
\begin{center}
\includegraphics[width=.31\textwidth]{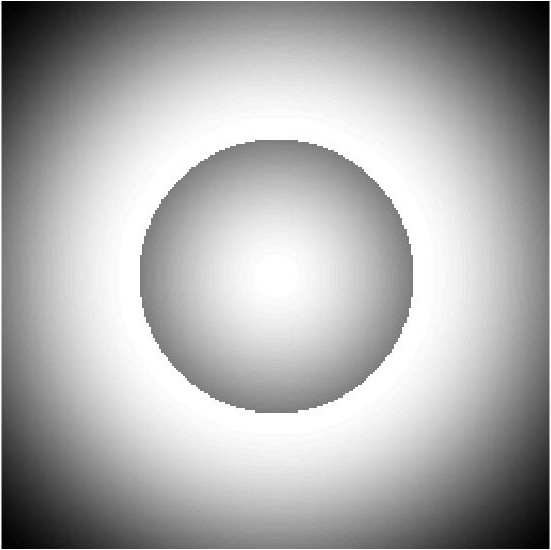}
\includegraphics[width=.31\textwidth]{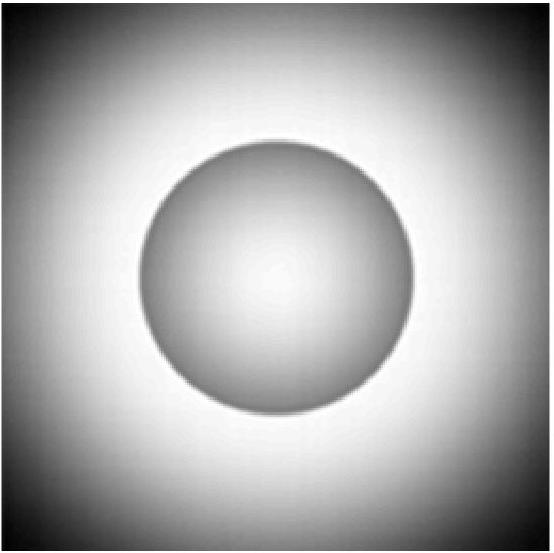}
\includegraphics[width=.34\textwidth]{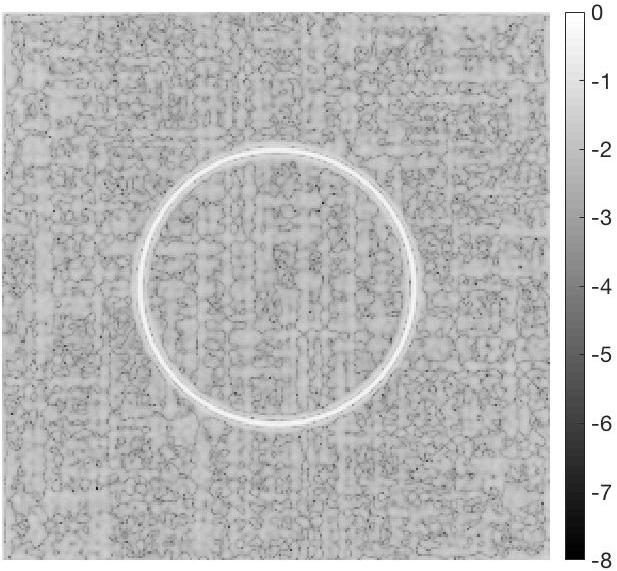}
\end{center}
\caption{(Left) $257\times257$ pixel function $f_3(x,y)$; (Middle) reconstruction via Algorithm \ref{alg:irl12d} with $m=2$ from $257\times257$ jittered Fourier coefficients on $257\times257$ grid points; (Right) pointwise error plotted on a logarithmic scale. The algorithm parameters used are $\rho=.01$, $\epsilon=.9$, and $\ell_{max}=5$.}
\label{fig:f3irl1}
\end{figure}

\begin{figure}[h!]
\begin{center}
\includegraphics[width=.35\textwidth]{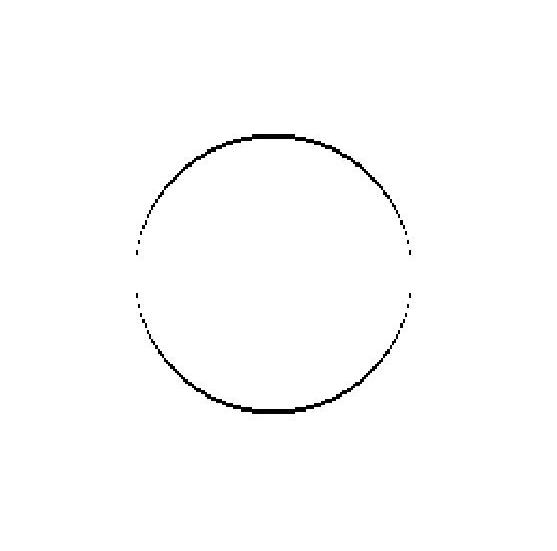}
\includegraphics[width=.35\textwidth]{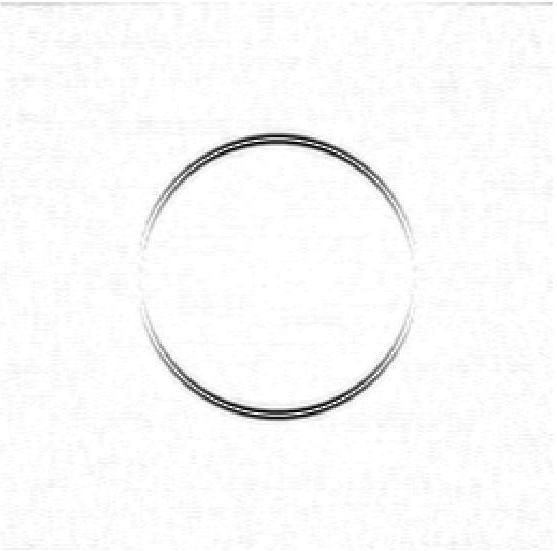}
\end{center}
\caption{(Left) Ideal weighting matrix for the $y$-direction edges computed from (\ref{eq:weights2d}) using the exact $257\times257$ two-dimensional function $f_3(x,y)$; (Right) final weighting matrix for the $y$-direction edges produced by Algorithm \ref{alg:irl12d}. The minimum weight of $0$ is indicated by black while maximum weight of $\frac{1}{\epsilon}\approx 1.11$ is indicated by white, while gray indicates a weight in between $0$ and $\frac1\epsilon$.}
\label{fig:f3weights}
\end{figure}

A two-dimensional example of this phenomenon can be seen in Figure \ref{fig:f3irl1}, which shows a reconstruction via Algorithm \ref{alg:irl12d} of the $257\times257$ pixel function $f_3(x,y)$. Figure \ref{fig:f3weights} elucidates the cause of these inaccuracies. The left image shows the ideal weights $w_{i,j}$ as in (\ref{eq:weights2d}). By ideal, we mean computed using (\ref{eq:weights2d}) directly from the exact function $f_3(x,y)$. The goal of Algorithm \ref{alg:irl12d} is to converge to these weights, since that would mean the algorithm generated $\mathbf{f}^{(\ell)}$ very close to the true $f_3(x,y)$. On the right we display the actual weights $w_{i,j}$ computed by Algorithm \ref{alg:irl12d}. The black or gray points in the right weighting matrix different from those obtained by the ideal left weighting matrix are being identified by Algorithm \ref{alg:irl12d} as either a jump or variation beyond the resolution of the problem.  Since there are no jumps outside of the black area in this function, this weighting matrix shows that the algorithm is falsely identifying many jumps. In this context, that means that this algorithm is regularizing less in areas it shouldn't be, which leads to an overall less accurate reconstruction.

In what follows we demonstrate how Algorithms \ref{alg:irl1} and \ref{alg:irl12d} can be improved upon in terms of accuracy, simplicity, efficiency and robustness.

\section{Edge-adaptive $\ell_2$ regularization} \label{EAR}

As discussed in \cite{candes2008enhancing}, without prior information about the non-zero elements in a sparse signal (or similarly the locations of edges in an image), it is effective to choose the $\ell_1$ regularization weights iteratively. However, when starting with Fourier data the weighting scheme adopted by Algorithms \ref{alg:irl1} and \ref{alg:irl12d} is likely not the most direct way to penalize non-zero locations in the edge domain. Therefore we take a more direct approach. Specifically we locate the edges directly from the Fourier data, as in \cite{gelb2017detecting}, and then construct the regularization weights to be zero-valued anywhere an edge is detected and non-zero at all other points. Unlike iterative reweighting, which requires multiple $\ell_1$ minimizations, there are only two minimization steps in our new method.  First, we perform an $\ell_1$ regularization based edge detection to determine where the support is in the sparsity (edge) domain. We then create a mask, i.e.~a weighting matrix, based on these regions. This mask allows us to target $\ell_2$ regularization only to non-edge regions of the function, which is the second minimization step. In this way, our method uses regularization on regions that are assumed to be zero.  Therefore the usual compressed sensing arguments for using the $\ell_1$ norm as a surrogate for the $\ell_0$ norm are not needed.  In particular, it is just as appropriate to use the $\ell_2$ norm to minimize something that is supposed to be zero, and it is much more cost efficient than using the $\ell_1$ norm.  We note that the method relies solely on the fidelity term in the (localized) support regions.

\subsection{Edge detection from non-uniform Fourier data}
\label{sec:edgedetection}
The edge adaptive $\ell_2$ regularization image reconstruction technique depends heavily on the selection of a weighting mask, which is explicitly determined by the edges recovered from the given non-uniform Fourier data.  While there have been a number of algorithms designed to extract edges from (non-uniform) Fourier data, we will use the method introduced in \cite{gelb2017detecting}.  It is briefly described below.

Let us first consider a one-dimensional piecewise smooth function $f:[-1,1]\rightarrow\mathbb{R}$. We define the jump function, $[f]$, as the difference between the left- and right-hand limits of the function:
\begin{align}\label{jumpdef}
[f](x) &= f(x^+)-f(x^-).
\end{align}
In smooth regions, $[f](x)=0$. At a discontinuity, $[f](x)$ is the value of the jump. Suppose we are given $2J+1$ grid points, $x_j = \frac{j}{J}$, $j=-J,\ldots,J$. Assuming that the discontinuities of $f$ are separated such that there is at most a single jump per cell, $I_j=[x_j,x_{j+1})$, we can write
\begin{align}
\label{eq:jumpdelta}
[f](x) & = \sum_{j=-J}^{J-1}[f](x_j)\delta_{x_j}(x).
\end{align}
where the coefficients $[f](x_j)$ is the value of the jump that occurs within the cell $I_j$ and $\delta_{x_j}(x)$ is the indicator function with 
$$\delta_{x_j}(x) =  \left\{
\begin{array}{lr}
1 & \text{if $x = x_j$}\\
0 & \text{otherwise.}
\end{array}
\right.
$$
The concentration factor (CF) edge detection method, \cite{gelb2008detection,gelb1999detection,gelb2000detection,gelb2006adaptive}, approximates (\ref{eq:jumpdelta}) from the first $2M+1$ uniform Fourier coefficients given in (\ref{eq:fourierdata}) where $\lambda_k=k$ as
\begin{align}
\label{eq:cf}
S_M^\sigma[f](x)& = i\sum_{k=-M}^{M}\hat{f}(k)\text{sgn}(k)\sigma(k) e^{\pi ikx}.
\end{align}
Here ${\bf \sigma} = \sigma(k)_{k = -M}^M$, coined the concentration factor in \cite{gelb1999detection}, satisfies certain admissibility conditions.  The convergence of (\ref{eq:cf}) depends on the particular choice of ${\bf \sigma}$.

The CF edge detection method cannot be extended directly to non-uniform Fourier coefficients because $\{e^{\pi i\lambda_k x}\}_{k=-M}^M$ is not an orthogonal basis. It is also not as effective when bands of data may be missing or corrupted, \cite{viswanathan2012iterative}. Therefore, as in \cite{gelb2017detecting} (see Algorithm 4), we approximate $[f]$ as the solution to the optimization problem
\begin{eqnarray}
\label{eq:edge}
\mathbf{g}^* &=& \arg\min_{\bf g} \left(\lvert \lvert  {\bf A} {\bf g} - diag({\bf \sigma}){\bf \hat f} \rvert \rvert_2 +\mu\lvert \lvert  {\bf g} \rvert \rvert_1\right).
\end{eqnarray}
That is, we fit the given Fourier data (here $diag({\bf \sigma}){\bf {\hat f}}$) using a forward NFFT operator ${\bf A}$ of the jump function vector ${\bf g} = \{[f](x_j)\}_{j = -J}^J$ and regularize with the sparsity of the jump function vector ${\bf g}$, with $\mu>0$ being the regularization parameter.  

As is discussed in \cite{gelb2011detection,gelb2017detecting,gelb1999detection},  one way to develop the CF edge detection method is to first observe that for a set of jump discontinuities $\{\xi_l\}_{l = 1}^L$, we have the first order approximation 
\begin{equation}
\label{eq:rampapproxoff}
f(x) \approx \sum_{l = 1}^L a_l r_{\xi_l}(x),
\end{equation}
where 
\begin{eqnarray}
\label{eq:ramp}
r(x) =  \left\{
\begin{array}{lr}
-\frac{x+1}{2} & \text{if $x \in [-1,0]$}\\
-\frac{x-1}{2} & \text{if $x \in (0,-1]$},
\end{array}
\right.
\end{eqnarray}
and $r_\xi(x) = r(x-\xi)$ for $\xi \in (-1,1)$.   Here $a_l$ is the corresponding jump value for $\xi_l$. While (\ref{eq:rampapproxoff}) is not a very good approximation of $f(x)$, it is perfectly reasonable to use to compute $[f](x)$.  Specifically, from (\ref{eq:rampapproxoff}) we have 
\begin{equation}
\label{eq:jumprampapprox}
[f](x) \approx \sum_{l = 1}^L a_l [r_{\xi_l}](x).
\end{equation}
This yields the $j^{th}$ row of ${\bf A}$ in (\ref{eq:edge}) as $e^{-i\pi x_j{\bf k}}{\bf \hat r}$ where ${\bf \hat r}$ are the Fourier coefficients of (\ref{eq:ramp}). However, as stated above, to improve efficiency we use the NFFT algorithm, \cite{greengard2004accelerating,lee2005type}. The CF vector ${\bf \sigma}$ is dependent on how $ \delta_{x_j}$ in (\ref{eq:jumpdelta}) is regularized, which is necessary since $\delta_{x_j}(x)$ has non-trivial values on a set of measure zero and therefore does not have a non-trivial Fourier expansion. As discussed in \cite{gelb2017detecting}, the concentration factors can be determined as
\begin{align}
{\bf \sigma}(k) &= \frac{\hat{h}_{x_j}(\lambda_k)}{\hat{r}(\lambda_k)},\quad{k = -M,\cdots,M,}
\end{align}
where $h_{x_j}(x) \approx \delta_{x_j}(x)$ and $\hat{h}_{x_j}(\lambda_k)$ are the corresponding Fourier coefficients at each $\lambda_k$.  For simplicity, in our experiments we choose each element ${\bf \sigma}(k) = \frac{2i\pi{ \lambda_k}}{2M+1}$ corresponding to regularization 
\begin{align}\label{eq:h}
h_{x_j}(x) &= \frac{2}{2M+1}\left(2\frac{\sin(\lambda_N\pi (x-x_j))}{\pi (x-x_j)} - 1\right).
\end{align}
We note that other options may provide better convergence in some examples. In particular, the Gaussian function
\begin{align}
h_{x_j}(x) = \exp\left(-5\left(\frac{x-x_j}{0.7}\right)^2\right)
\end{align}
can provide smoothing in the presence of noise. In our testing with jittered data, however, we found no difference in performance, and in general used (\ref{eq:h}). We refer readers to \cite{gelb2017detecting} for a detailed analysis of the terms in (\ref{eq:edge}).  Figure \ref{fig:1Dedge} demonstrates the use of (\ref{eq:edge}) on $f_1(x)$ and $f_2(x)$.

\begin{figure}[h!]
\begin{center}
\includegraphics[width=.32\textwidth]{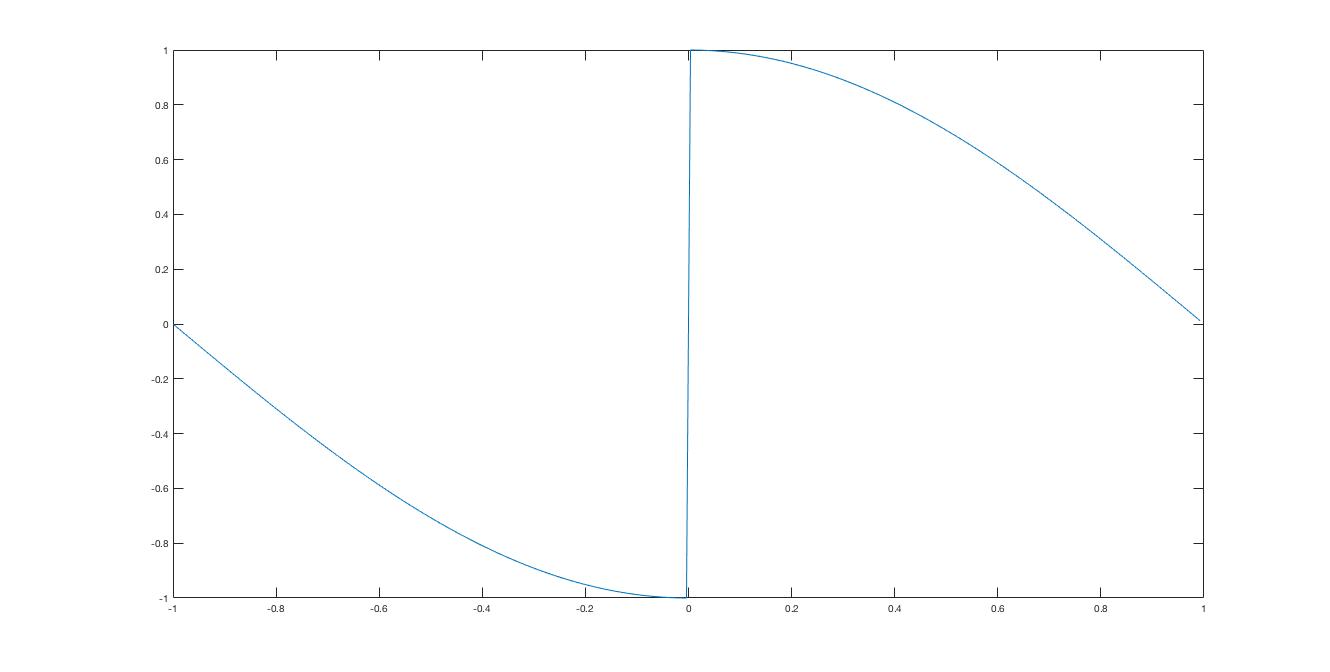}
\includegraphics[width=.32\textwidth]{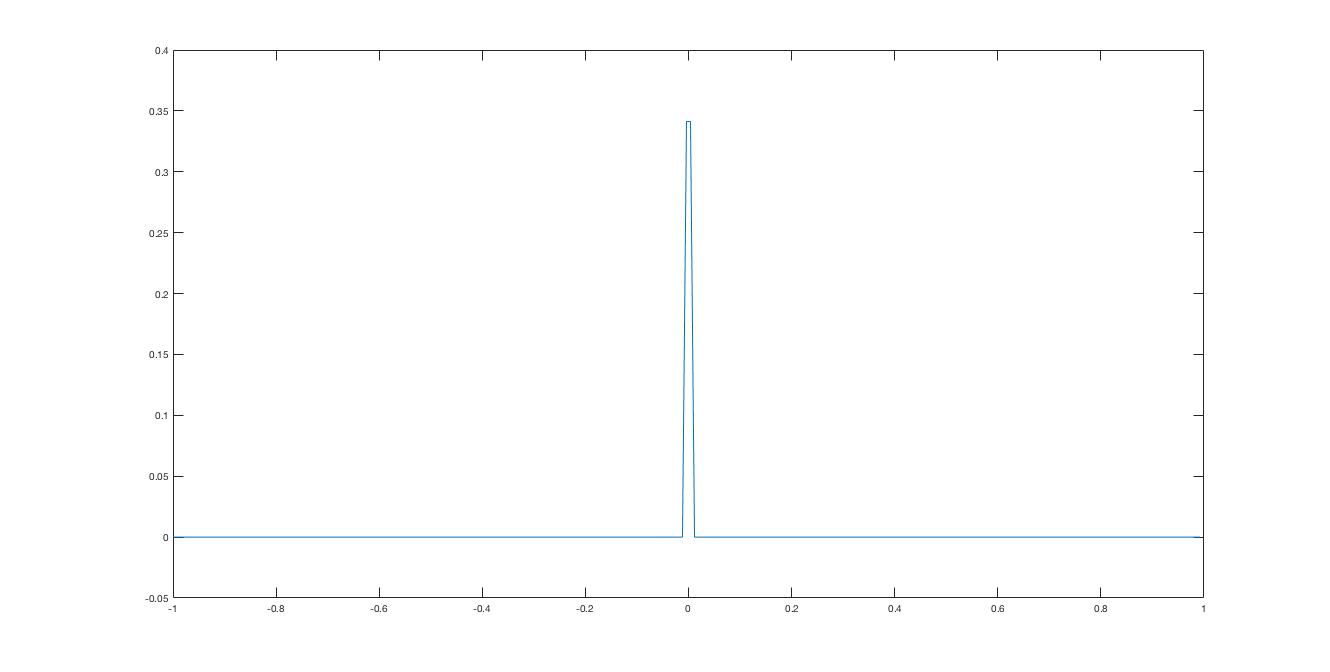}
\includegraphics[width=.32\textwidth]{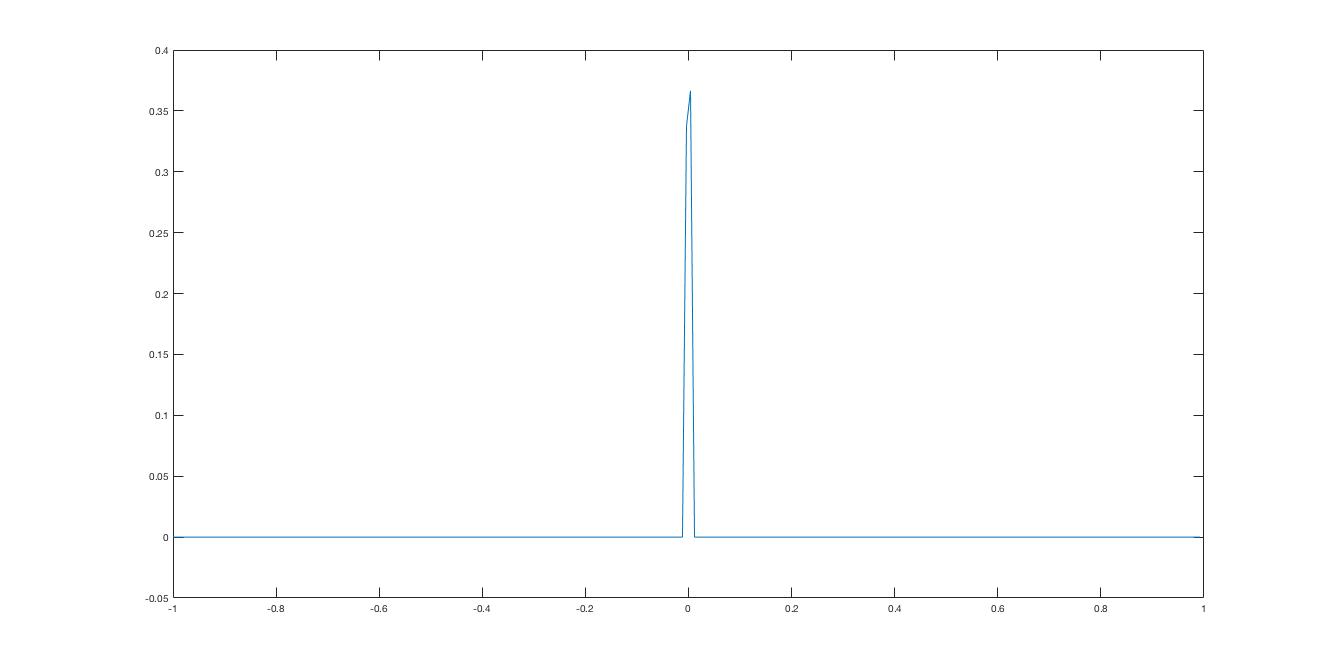}
\includegraphics[width=.32\textwidth]{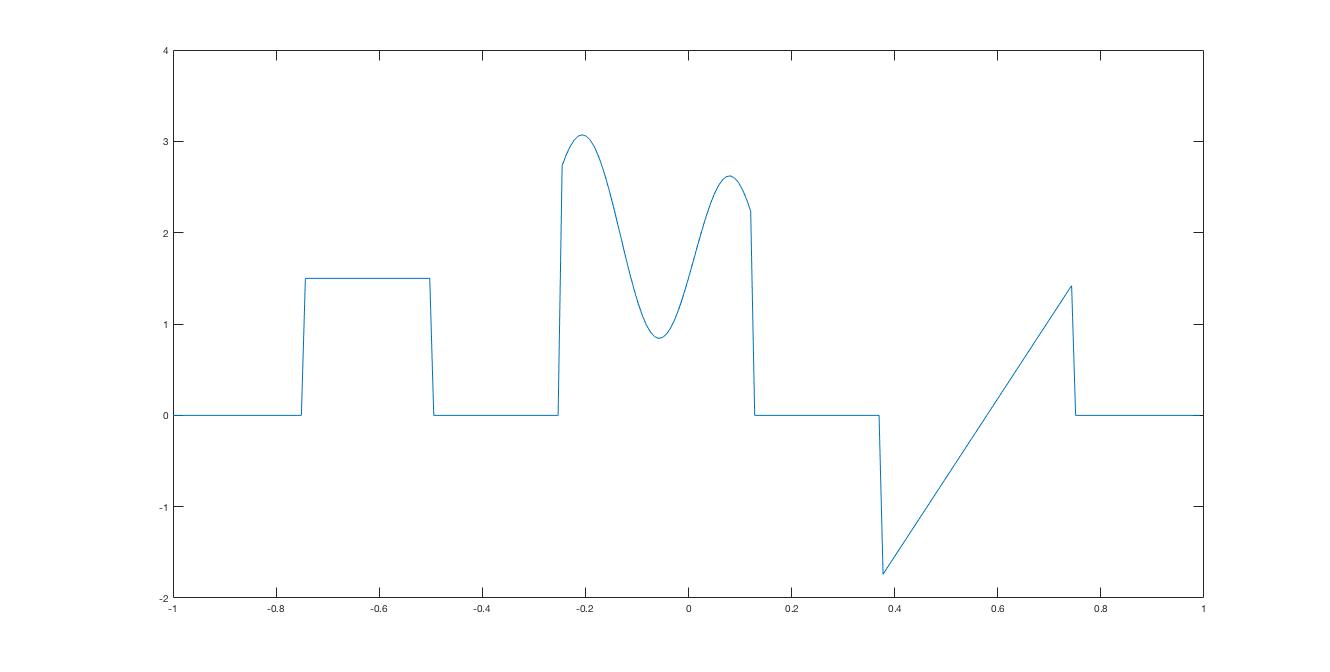}
\includegraphics[width=.32\textwidth]{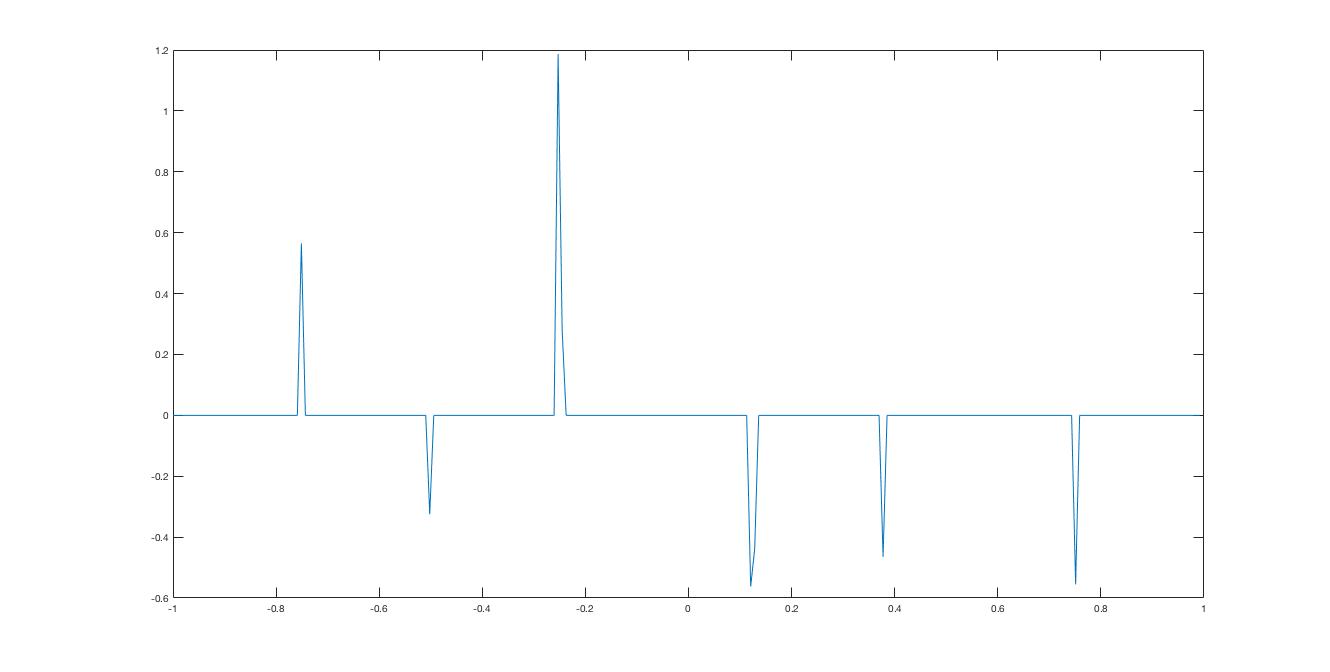}
\includegraphics[width=.32\textwidth]{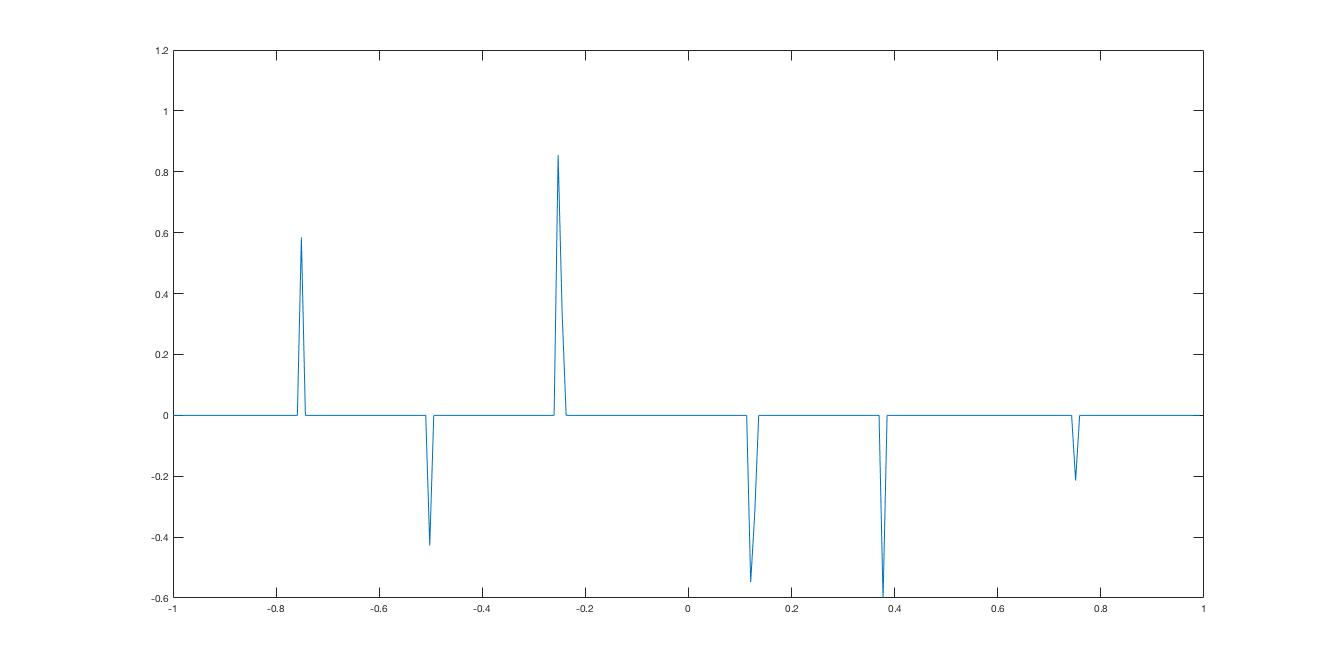}
\end{center}
\caption{(Left) $f_1(x)$ and $f_2(x)$; (Middle) Jump function approximations using (\ref{eq:edge}); (Right) Jump function approximations with additive noise starting from (\ref{eq:noisydata}) with SNR $=20$ dB. Here we use $257$ reconstruction points, $257$ jittered Fourier modes, and regularization parameter $\mu=1$.}
\label{fig:1Dedge}
\end{figure}

In two dimensions, the jump functions in the $x$ and $y$ directions may be approximated by the respective solutions to the optimization problems
\begin{eqnarray}
\label{eq:edge2d}
\mathbf{g}^*_x & = & \arg\min_{\bf g}  \lvert \lvert  {\bf A} {\bf g} - {diag({\bf \sigma_x}){\bf \hat f}} \rvert \rvert_2 +\mu\lvert \lvert  {\bf g} \rvert \rvert_1
 \nonumber \\
\mathbf{g}^*_y & = & \arg\min_{\bf g} \lvert \lvert  {\bf g}{\bf A^T}  - {diag({\bf \sigma_y}){\bf \hat f}} \rvert \rvert_2 +\mu\lvert \lvert  {\bf g} \rvert \rvert_1,
\end{eqnarray}
where $\sigma_x(\mathbf{k})=\frac{2i\pi\lambda_{k_1}}{2M+1}$ and $\sigma_y(\mathbf{k}) = \frac{2i\pi\lambda_{k_2}}{2M+1}$, $k_1,k_2 = -M,\cdots,M.$   We combine these $x$ and $y$ direction edge approximations into a single edge map by
\begin{eqnarray}
\label{eq:edge2dcombo}
\mathbf{g}^*(x_t,y_s) & = & \max\left\{\left|\mathbf{g}^*_x(x_t,y_s)\right|,\left|\mathbf{g}^*_y(x_t,y_s)\right|\right\}
\end{eqnarray}
for each point $(x_t,y_s)$ in the grid. Figure \ref{fig:edgecos2d} shows the results for approximating $[f_3](x,y)$ in (\ref{eq:f2D}).
We note that the definition of jump value in (\ref{jumpdef}) does not carry over to two dimensions. Indeed the approximation in (\ref{eq:edge2dcombo}) can be replaced as the average values from (\ref{eq:edge2d}) or something else entirely.  The actual value of the jump is not critical for our purposes since we need only identify where the jump function is non-zero.
\begin{figure}[h!]
\begin{center}
\includegraphics[width=.35\textwidth]{2Dhill.jpg}
\includegraphics[width=.35\textwidth]{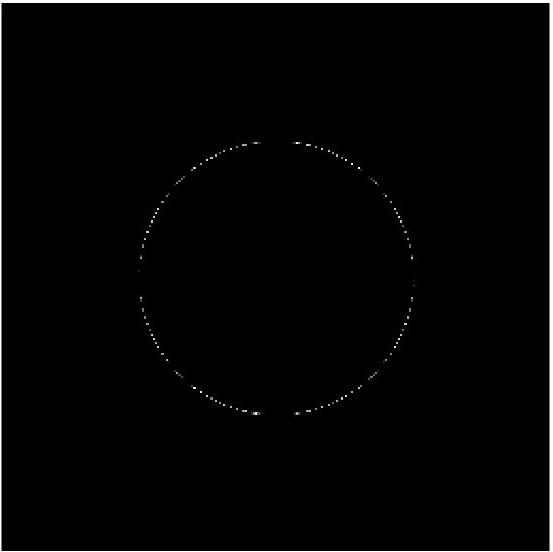}
\end{center}
\caption{Edge detection (right) using (\ref{eq:edge2d}) and (\ref{eq:edge2dcombo}) for $f_3(x,y)$ (left). Here we use $257\times257$ jittered Fourier modes, $257\times257$ reconstruction points, and regularization parameter $\mu=1$.}
\label{fig:edgecos2d}
\end{figure}

Edge detection in and of itself can be an important tool in identifying physical structures in images or signals. In MRI, edge detection helps tissue boundary identification. In SAR, it can improve target identification. Unlike other algorithms, a potentially useful byproduct of our new algorithm is that we can also output an accurate {\em edge map} that we compute en route to the reconstruction. In particular, it can act as a cross-validation for the reconstruction.  For the purposes of our reconstruction, we need only a binary edge map. In one dimension, this is easily generated as
\begin{eqnarray}
\label{eq:binaryedge1d}
{\bf y}_j =  \left\{
\begin{array}{lr}
1 & \text{if $|{\bf g}^*_j| > \tau$}\\
0 & \text{otherwise,}\
\end{array}
\right.
\hspace{.2in}  j=-J,\ldots,J.
\end{eqnarray}
Here ${{\bf g}^*}$ is determined from (\ref{eq:edge}). The resolution-dependent threshold $\tau$ is typically inversely proportional to the number of grid points, meaning that with better resolution we should be able to detect jumps of smaller magnitude. When noise is present, the threshold is also dependent on the SNR. In two dimensions, we generate the binary edge maps as
\begin{eqnarray}
\label{eq:binaryedge2d}
{\bf x}_{i,j} =  \left\{
\begin{array}{lr}
1 & \text{if $|({\bf g}^*_x)_{i,j}| > \tau$}\\
0 & \text{otherwise,}\
\end{array}
\right.
\hspace{.2in}  i,j=-J,\ldots,J.
\end{eqnarray}
and
\begin{eqnarray}
\label{eq:binaryedge2d}
{\bf y}_{i,j} =  \left\{
\begin{array}{lr}
1 & \text{if $|({\bf g}^*_y)_{i,j}| > \tau$}\\
0 & \text{otherwise,}\
\end{array}
\right.
\hspace{.2in}  i,j=-J,\ldots,J.
\end{eqnarray}
where $\mathbf{g}^*_x$ and $\mathbf{g}^*_y$ are determined from (\ref{eq:edge2d}). We note that another potential benefit of utilizing the edge map is that it only ever needs to be computed once per image. This means that if we have an edge map from another experiment, no matter how it was obtained, it can be used here too.

As will be described below, edge detection and the generation of the binary edge map are critical in determining the weighting mask for the regularization term,  and is even more important when the given data are noisy.  In particular, edge information enables us to adapt the regularization parameter to weight relatively lightly on fidelity in non-edge areas and heavily on fidelity in edge areas where we are in general more confident of non-zero values in the sparsity domain.

\subsection{Edge-adaptive $\ell_2$ regularization image reconstruction algorithm}

Once the binary edge map in (\ref{eq:binaryedge1d}) or (\ref{eq:binaryedge2d}) is formed, the next step of the edge-adaptive $\ell_2$ regularization method is to create a weighting matrix, or mask.  This process is detailed in Algorithm \ref{alg:mask} for the one-dimensional case and Algorithm \ref{alg:mask2d} for the two-dimensional case.  When the PA transform is used in the regularization term for the reconstruction procedure, the corresponding regularization mask must include the stencil corresponding to the degree of the PA transform used. This is because the PA transform forms an oscillatory response in the stencil surrounding the jump. For example, when using $L^3$ in (\ref{eq:Lm}), there is an oscillatory response in $4$ points surrounding a detected jump (including the pixel value associated with the jump).  We note that the stencil size of any weighting mask would directly correspond to the degree of the corresponding sparsifying transform operator (e.g.~wavelets).
\begin{algorithm}[h!]
\caption{Mask creation in one dimension}
\label{alg:mask}
\begin{algorithmic}[1]
\STATE Starting from Fourier data as in (\ref{eq:noisydata}), reconstruct the jump function, $[f]$, using equation (\ref{eq:edge}) as $\mathbf{g}^*$.

\STATE For each index $j=-J,\ldots,J$ such that $|\mathbf{g}^*_{j}|>\tau$, set $\mathbf{y}_{j}=1$. Else, $\mathbf{y}_{j}=0$.

\STATE For each index $j=-J,\ldots,J$ such that $|(L^m \mathbf{y})_{j}|>\tau$, set $\mathbf{z}_{j}=0$. Else, $\mathbf{z}_{j}=1$. The mask is $M = \text{diag}(\mathbf{z})$.
\end{algorithmic}
\end{algorithm}

\begin{algorithm}[h!]
\caption{Mask creation in two dimensions}
\label{alg:mask2d}
\begin{algorithmic}[1]
\STATE Starting from Fourier data as in (\ref{eq:fourierdata2d}), reconstruct the jump functions in the $x$ and $y$ directions using equations (\ref{eq:edge2d}) as $\mathbf{g}_x^*$ and $\mathbf{g}_y^*$ and combine into a single edge map, $\mathbf{g}^*$, using (\ref{eq:edge2dcombo}).

\STATE For each index $i,j=-J,\ldots,J$ such that $|(\mathbf{g}^*_x)_{i,j}|>\tau$, set $\mathbf{x}_{i,j}=1$. Else, $\mathbf{x}_{i,j}=0$.

\STATE For each index $i,j=-J,\ldots,J$ such that $|(\mathbf{g}^*_y)_{i,j}|>\tau$, set $\mathbf{y}_{i,j}=1$. Else, $\mathbf{y}_{i,j}=0$.

\STATE For each index $i,j=-J,\ldots,J$ such that $|(L^m \mathbf{x})_{i,j}|>\tau$, set $M^x_{i,j}=0$. Else, $M^y_{i,j}=1$. The $x$ direction mask is the matrix $M^x$.

\STATE For each index $i,j=-J,\ldots,J$ such that $|(\mathbf{y}(L^m)^T)_{i,j}|>\tau$, set $M^y_{i,j}=0$. Else, $M^y_{i,j}=1$. The $y$ direction mask is the matrix $M^y$.
\end{algorithmic}
\end{algorithm}

Observe that unlike the iterative weights described in Section \ref{IR}, the weighting masks generated by Algorithms \ref{alg:mask} and \ref{alg:mask2d}  are binary. There are three advantages to our technique:  (i)  accuracy is improved since in general we do not falsely identify jumps; (ii) our method is more efficient since we do not have to perform expensive iterations to locate the edges; and (iii) we improve the efficiency even further since we do not need to use $\ell_1$ regularization in the reconstruction step.  This is because we have reframed the optimization problem in (\ref{eq:HOTV}) as:
\begin{equation}
\arg\min_\mathbf{g} ||ML^m {\bf g}||_2
\hspace{.2in} \mbox{subject to }
||{\mathcal F}_{N}{\bf g} - {\bf \hat f}||_2 < \delta
\label{eq:newl2}
\end{equation}
where $\delta>0$ is a threshold on the fidelity term. Observe that minimizing $||ML^m{\bf g}||_2$ in (\ref{eq:newl2}) is equivalent to setting up the usual sparsity constraint, which  requires $L^m g$  to have only a few non-zero values, or more precisely, values above a chosen threshold.  Algorithm \ref{alg:eal2} demonstrates how the constrained optimization problem (\ref{eq:newl2}) can be solved by converting it into an equivalent unconstrained optimization problem. Algorithm \ref{alg:eal22d} details the algorithm for two-dimensional functions and images.

\begin{algorithm}[h!]
\caption{Edge-adaptive image reconstruction in one dimension}
\label{alg:eal2}
\begin{algorithmic}[1]

\STATE Construct the mask, $M$, using Algorithm \ref{alg:mask}.

\STATE The edge-adaptive $\ell_2$ regularization image reconstruction is the solution to the optimization problem,
\begin{eqnarray}
\label{eq:EAR}
 {\bf f}^* & = & \arg\min_{\bf g} \left( \lvert \lvert  {\mathcal F}_{N}{\bf g} - {\bf \hat f} \rvert \rvert^2_2 +  \lambda \lvert \lvert  ML^m{\bf g} \rvert \rvert^2_2\right),
\end{eqnarray}
where $\lambda>0$ is the regularization parameter. 

\end{algorithmic}
\end{algorithm}

We note that (\ref{eq:EAR}) has a closed form solution
\begin{eqnarray}
\label{eq:tikhsol}
 {\bf f}^*& = &  \left({\mathcal F}_{N}^T{\mathcal F}_{N} + \lambda(L^m)^TML^m\right)^{-1} {\mathcal F}_{N}^T{\bf \hat{f}}.
\end{eqnarray}
This closed form may be valuable in some contexts. As the size of the problem increases, however, the inversion in (\ref{eq:tikhsol}) becomes more computationally expensive, so in Section \ref{numerics} we take advantage of the conjugate gradient descent method \cite{golub2012matrix}.

\begin{algorithm}[h!]
\caption{Edge-adaptive image reconstruction in two dimensions}
\label{alg:eal22d}
\begin{algorithmic}[1]

\STATE Construct the mask, $M$, using Algorithm \ref{alg:mask2d}.

\STATE The edge-adaptive $\ell_2$ regularization image reconstruction is the solution to the optimization problem,
\begin{equation}\label{eq:EAR2d}
\begin{split}
 {\bf f}^* & = \arg\min_{\bf g} \left(\lvert \lvert  {\mathcal F}_{N}{\bf g} - {\bf \hat f} \rvert \rvert^2_2\right.\\
 & + \left. \lambda\left(\sum_{i=-J}^J\sum_{j=-J}^JM^x_{i,j}(L^m{\bf g})_{i,j}^2+\sum_{i=-J}^J\sum_{j=-J}^JM^y_{i,j}({\bf g}(L^m)^T)_{i,j}^2\right)\right),
\end{split}
\end{equation}
where $\lambda>0$ is the regularization parameter.

\end{algorithmic}
\end{algorithm}

\section{Numerical results}\label{numerics}

In the numerical experiments that follow we compare the edge-adaptive $\ell_2$ regularization image reconstruction given by Algorithms \ref{alg:eal2}  and  \ref{alg:eal22d}  to the iteratively reweighted method of Algorithm \ref{alg:irl1} and  \ref{alg:irl12d}. We use the Split Bregman method, \cite{goldstein2009split,yin2008bregman} to implement the minimization step in Algorithms \ref{alg:irl1} and \ref{alg:irl12d}. We follow the recommendation in \cite{candes2008enhancing} in choosing the parameter $\epsilon$ to be  slightly smaller than the expected nonzero magnitudes of $L^m\mathbf{f}$, since this will  provide the necessary stability to correct for inaccurate coefficient estimates while still improving upon the unweighted TV algorithm. We note that in \cite{chartrand2008iteratively} it is shown that updating $\epsilon$ in each iteration yields superior results for the problem of sparse signal recovery. However, since this is not applicable for functions with more variation, we did not consider this adaptive approach.  Algorithms \ref{alg:eal2} and \ref{alg:eal22d}, which only require $\ell_2$-regularized minimization, are performed using conjugate gradient descent, \cite{golub2012matrix}. In what follows we look at examples in both one and two dimensions and the results using these algorithms for different regularization parameter. We also vary the PA order $m$, add noise to the initial data, and limit the amount of initial data. In all cases we compare the accuracy and efficiency of each algorithm. Finally, we demonstrate the success of our new algorithm on synthetic aperture radar (SAR) data, \cite{dungan2010civilian}.

\subsection*{One-dimensional test case}
Figure \ref{fig:1Dcos} compares the results of Algorithm \ref{alg:irl1} and Algorithm \ref{alg:eal2} for $f_1(x)$ in (\ref{eq:f1}), where the acquired data are $257$ noise-free jittered Fourier samples given by (\ref{eq:fourierdata}). We computed the relative error,
\begin{equation}
\label{eq:rel_err}
RE = ||\mathbf{f}^*-\mathbf{f}||_2/||\mathbf{f}||_2,
\end{equation}
for each algorithm, resulting in $RE = .0446$ using Algorithm \ref{alg:irl1}  and $RE = .0155$ using Algorithm \ref{alg:eal2}. In addition to improving the overall accuracy, it is evident that due to the precise jump identification yielded using (\ref{eq:binaryedge1d}), there is improved resolution and reduced error in the neighborhood of the jump.

\begin{figure}[h!]
\begin{center}
\includegraphics[width=.40\textwidth]{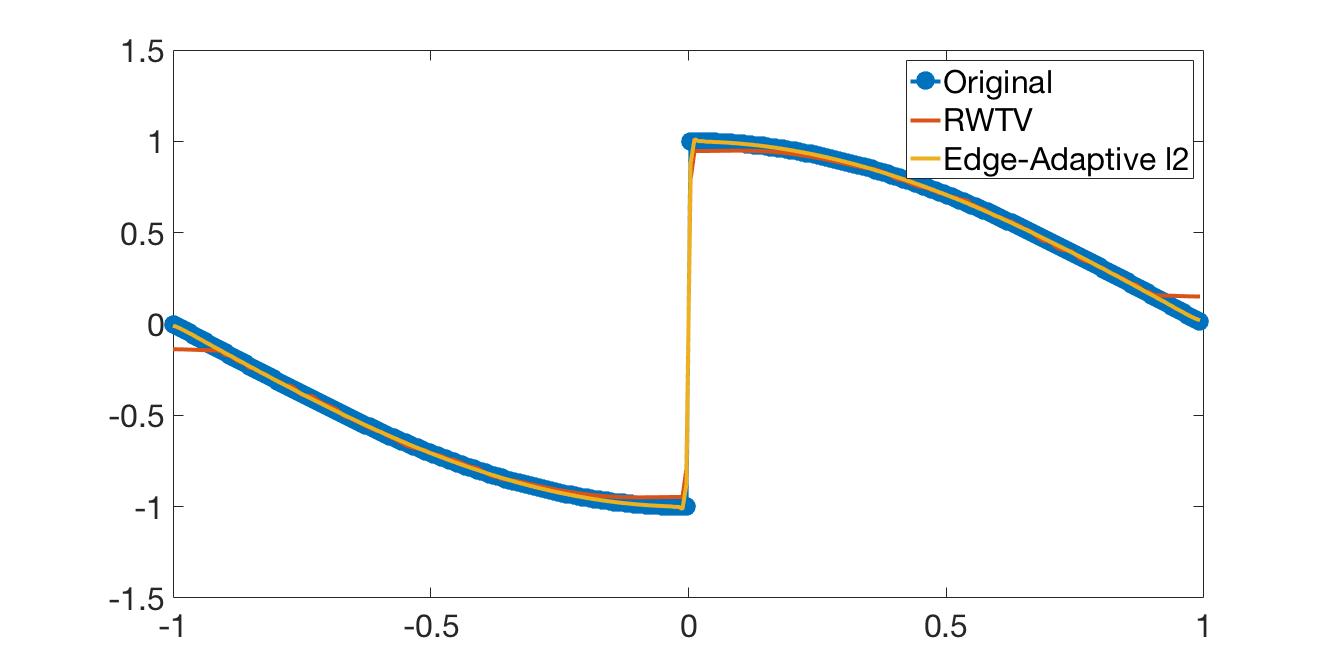}
\includegraphics[width=.40\textwidth]{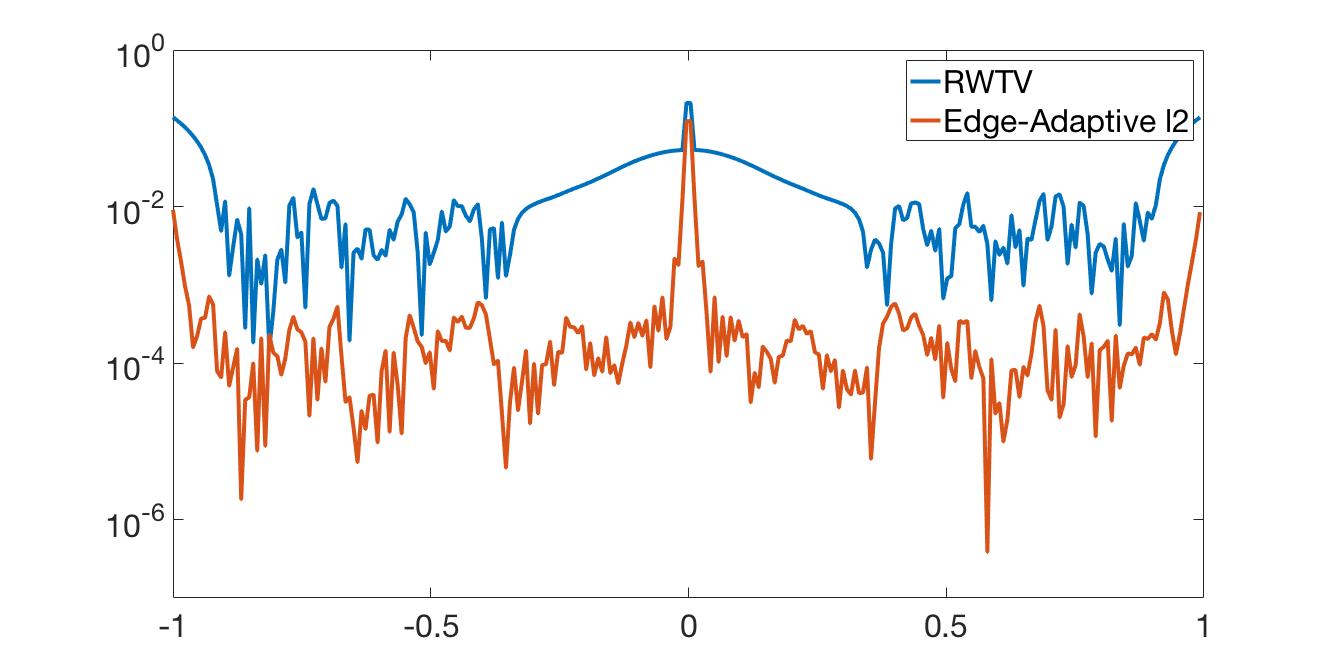}
\end{center}
\caption{(Left) Comparison of Algorithms \ref{alg:irl1} and \ref{alg:eal2} on $f_1(x)$ using PA order $m=1$ given $257$ jittered Fourier samples reconstructed on $257$ grid points; (Right) corresponding pointwise errors. For parameters, we use $\rho=1$, $\mu=1$, $\lambda=1$, $\epsilon=1.9$, $\ell_{max}=25$, and threshold $\tau=1/257$.}
\label{fig:1Dcos}
\end{figure}

\subsection*{Two-dimensional test case}

\begin{figure}[htbp]
\begin{center}
\includegraphics[width=.35\textwidth]{2Dhill_irl1_2TV.jpg}
\includegraphics[width=.35\textwidth]{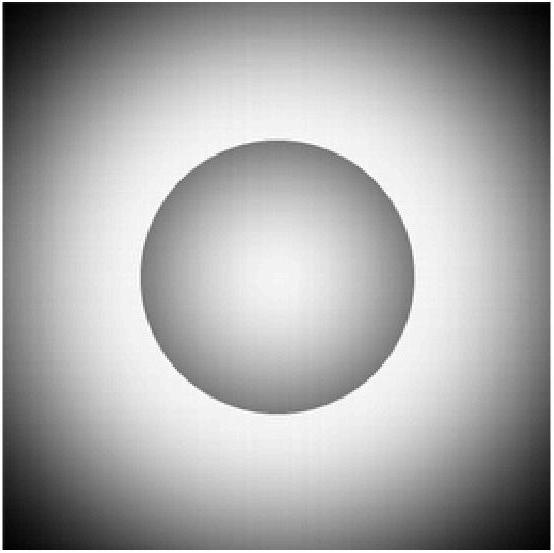}
\end{center}
\caption{Comparison of Algorithms \ref{alg:irl12d} (left) and \ref{alg:eal22d} (right) for reconstructing $f_3(x,y)$ from $257\times257$ noise-free Fourier modes on $257\times257$ grid points. For parameters, we use PA order $m=2$ due to the piecewise quadratic nature of the function, $\rho=.01$, $\epsilon=.9$, and $\ell_{max}=5$, $\mu=.1$, $\tau=.025$, and $\lambda=1$.}
\label{fig:recon_2d_cos_fxn}
\end{figure}

\begin{figure}[htbp]
\begin{center}
\includegraphics[width=.35\textwidth]{2Dhill_irl1_2TV_error.jpg}
\includegraphics[width=.35\textwidth]{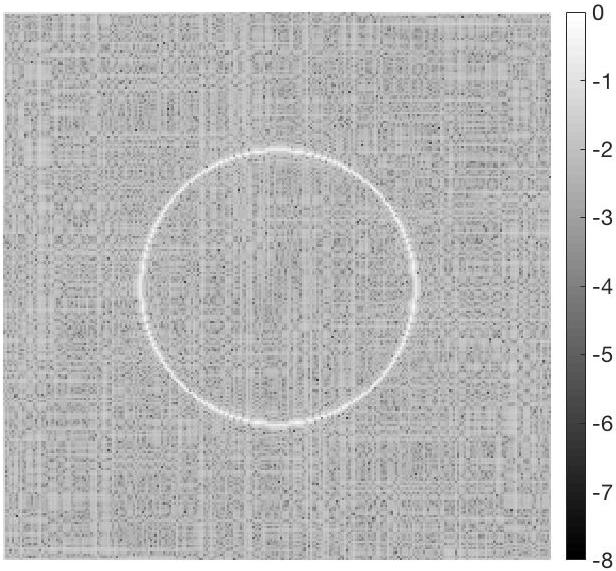}
\end{center}
\caption{Comparison of errors using Algorithms \ref{alg:irl12d} (left) and \ref{alg:eal22d} (right) for reconstructing $f_3(x,y)$, using the same parameters as Figure \ref{fig:recon_2d_cos_fxn}.}
\label{fig:recon_2d_cos_fxn_error}
\end{figure}

Similar results are obtained in two dimensions, as is confirmed in Figures \ref{fig:recon_2d_cos_fxn} and \ref{fig:recon_2d_cos_fxn_error}, which compare the results using Algorithms \ref{alg:irl12d} and \ref{alg:eal22d} for $f_3(x,y)$ given in (\ref{eq:f2D}). The data acquired are $257^2$ noise-free jittered Fourier samples given by (\ref{eq:fourierdata2d}). It is evident that Algorithm \ref{alg:eal22d} yields both better overall accuracy in terms of relative error, $RE = .0414$ for Algorithm \ref{alg:eal22d} versus $RE = .0616$ for Algorithm \ref{alg:irl12d}, as well as improved resolution near the edges of the image, exhibited by the smaller white ring in its error plot.

\subsection*{Robustness of regularization parameter}
Choosing the regularization parameter for the minimization step of optimization-based reconstruction methods, e.g.~as in (\ref{eq:HOTV}), is typically difficult and problem-dependent, \cite{osher2005iterative}, yet crucial to the success of the algorithm. Using the edge-adaptive $\ell_2$ method, we observe a robustness with respect to the choice of this parameter. This is shown in Figure \ref{fig:1Dlambda} (right), which displays the pointwise error plots comparing Algorithm \ref{alg:irl1} (RWTV) and Algorithm \ref{alg:eal2} for reconstructing $f_1(x)$ in (\ref{eq:f1}) for various values of $\lambda$.  Observe that our edge-adaptive $\ell_2$ method outperforms the RWTV reconstruction for a wide range of $\lambda$.  Such robustness is critical since in many applications reliable ground truth information is not available.   In terms of relative error, Algorithm \ref{alg:irl1} yielded $RE = .0446$, while even in the worst case, $\lambda = 100$, Algorithm \ref{alg:eal2} produced  $RE = .0478$. Moreover, it is evident that for all choices of $\lambda$, there is improved resolution using our algorithm in neighborhoods of the jump. These results are particularly impressive when compared with the robustness of Algorithm \ref{alg:irl1} with respect to the regularization parameter $\rho$ as seen in Figure \ref{fig:1Dlambda} (left). There we see that the accuracy varies strongly with $\rho$, in particular around the jump. Regardless of the choice of $\lambda$, Algorithm \ref{alg:eal2} outperformed Algorithm \ref{alg:irl1}, especially near the jump.

\begin{figure}[h!]
\begin{center}
\includegraphics[width=.49\textwidth]{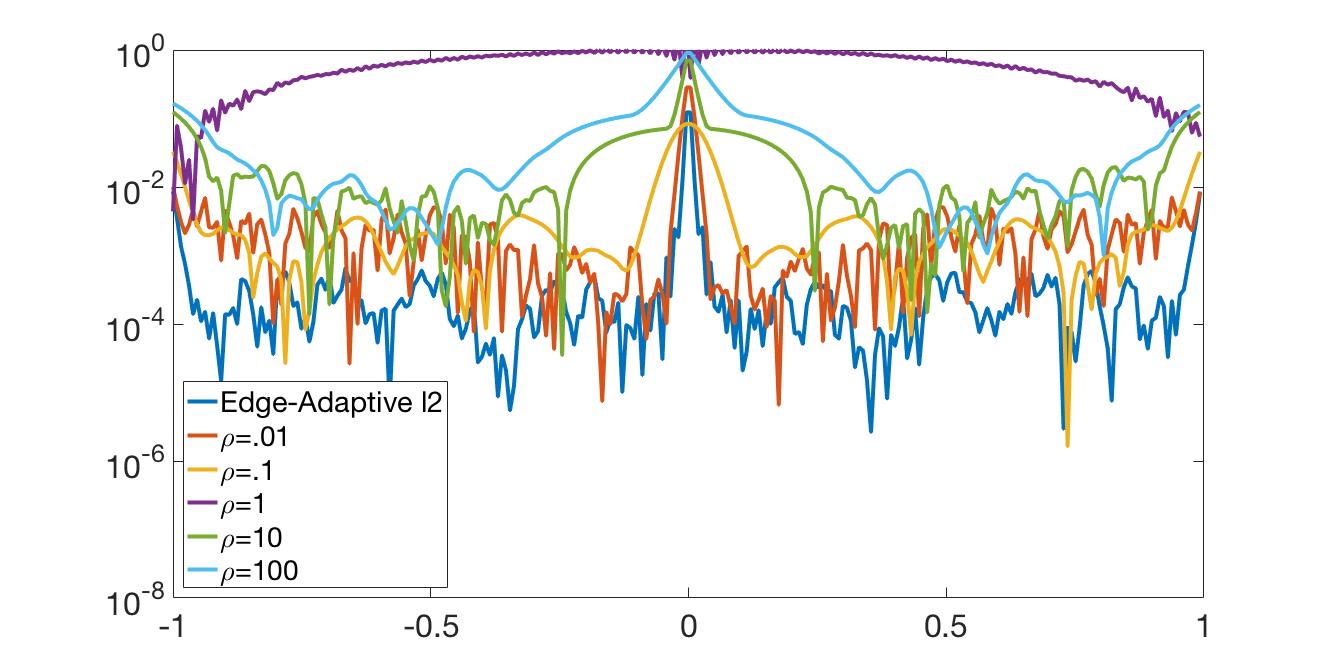}
\includegraphics[width=.49\textwidth]{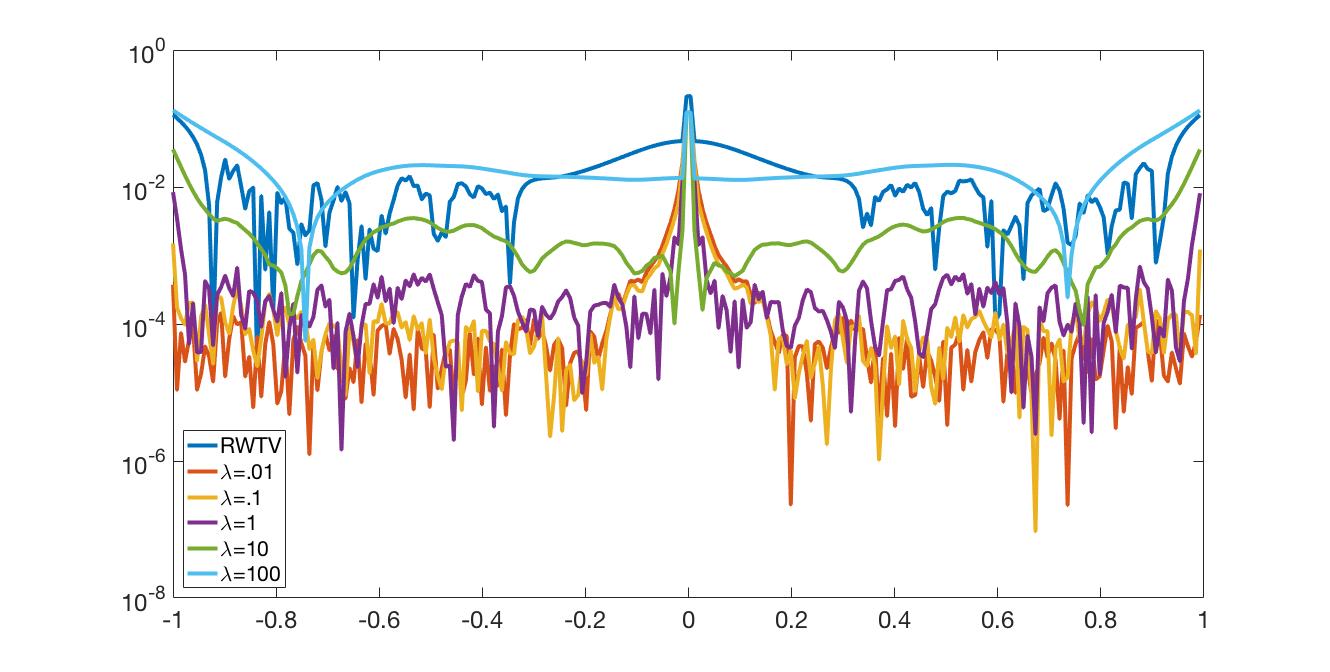}
\end{center}
\caption{Pointwise error plot comparisons between Algorithm \ref{alg:irl1} and Algorithm \ref{alg:eal2}  for $f_1(x)$ given $257$ jittered Fourier samples reconstructed on $257$ grid points. (Left) Parameters $m = 1$, $\ell_{max}=25$, $\epsilon=1.9$, $\tau=1/257$, $\lambda=1$, and vary $\rho=.01,.1,1,10,100$; (Right) Parameters $m = 1$, $\rho=1$, $\ell_{max}=25$, $\epsilon=1.9$, $\mu=1$, $\tau=1/257$, and vary $\lambda=.01,.1,1,10,100$.}
\label{fig:1Dlambda}
\end{figure}

\subsection*{Comparison of PA order}
\begin{figure}[htb]
\begin{center}
\includegraphics[width=.35\textwidth]{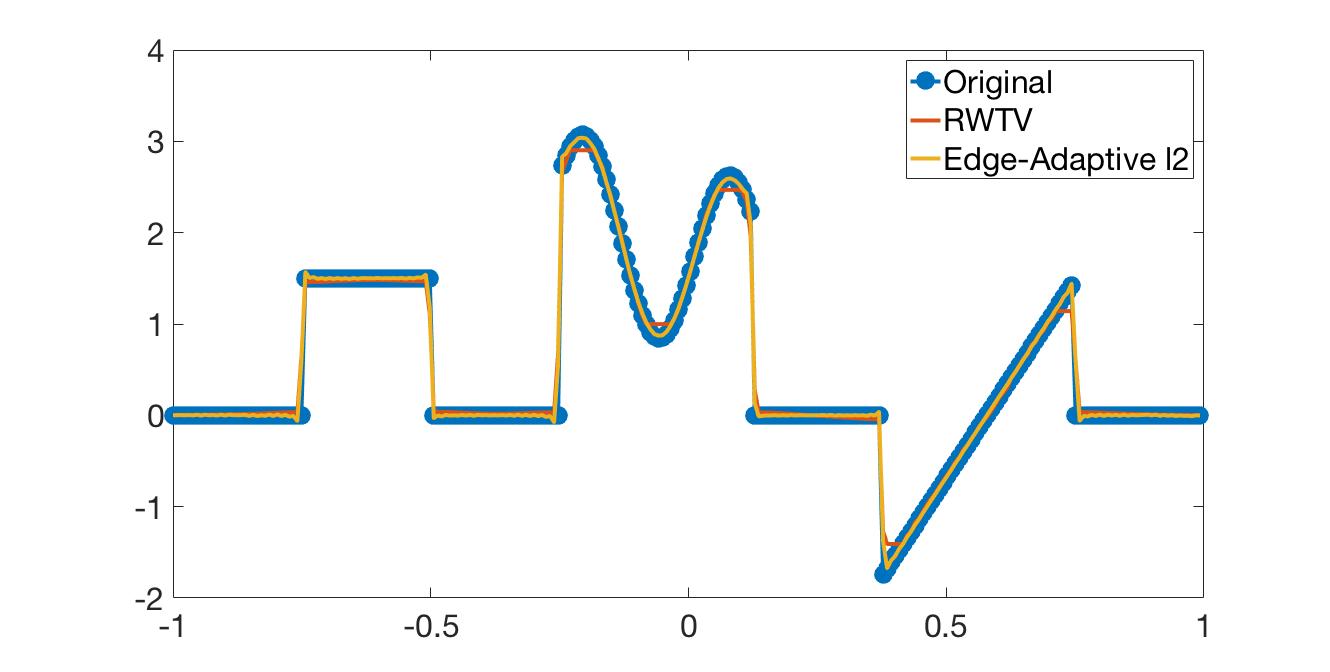}
\includegraphics[width=.35\textwidth]{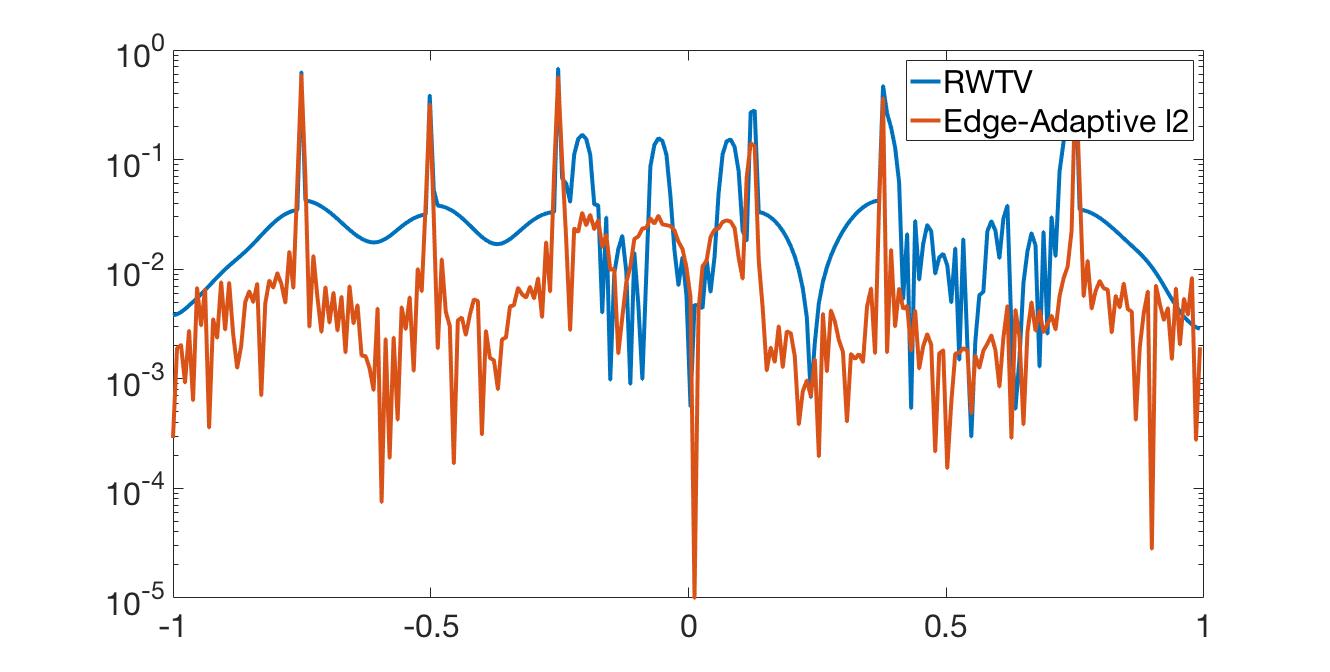}
\includegraphics[width=.35\textwidth]{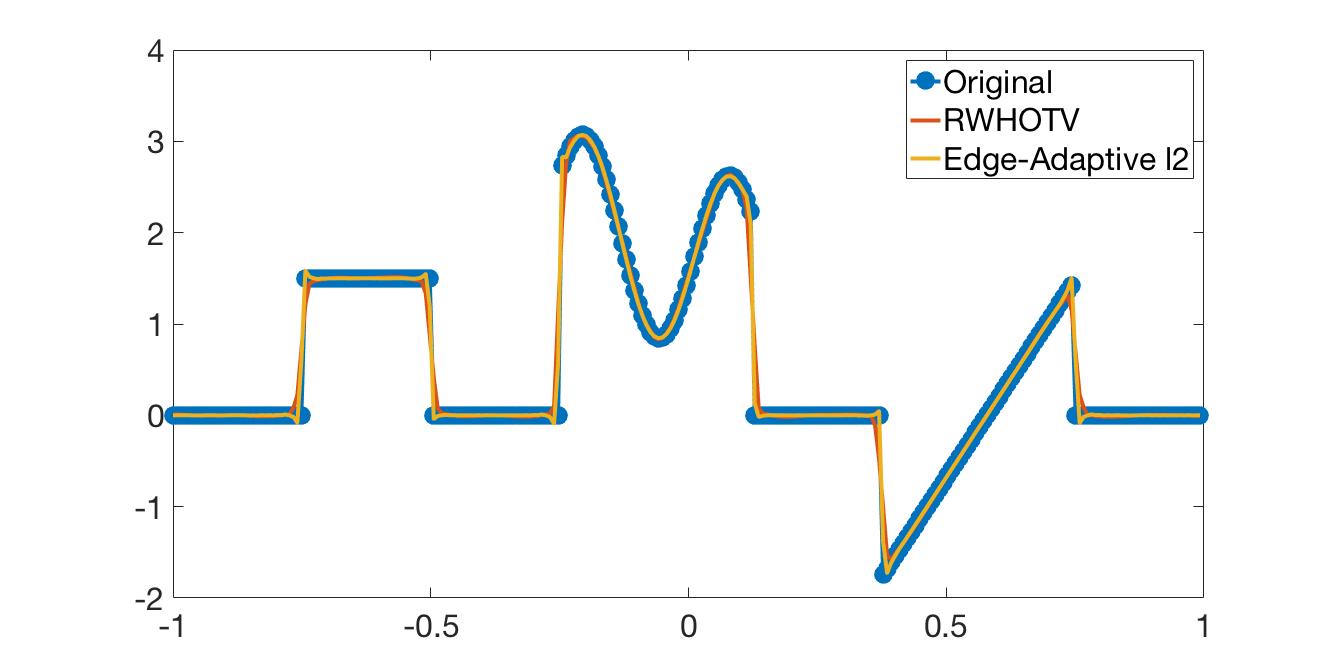} 
\includegraphics[width=.35\textwidth]{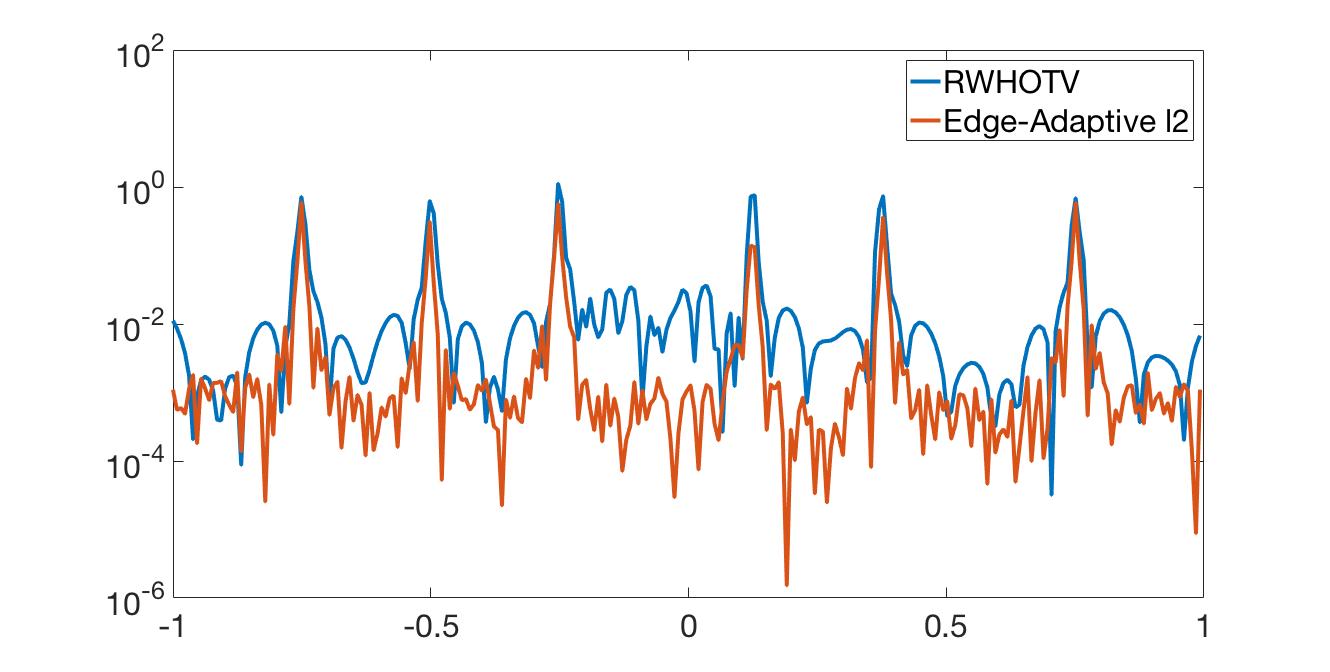}
\includegraphics[width=.35\textwidth]{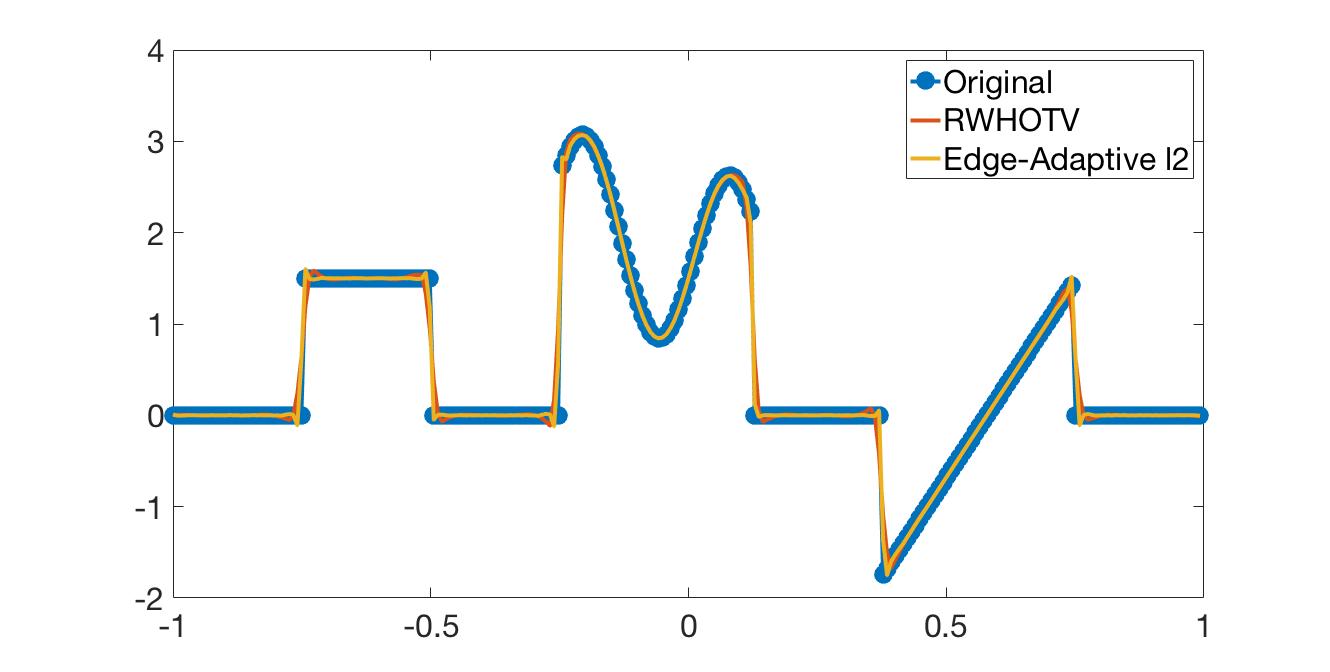} 
\includegraphics[width=.35\textwidth]{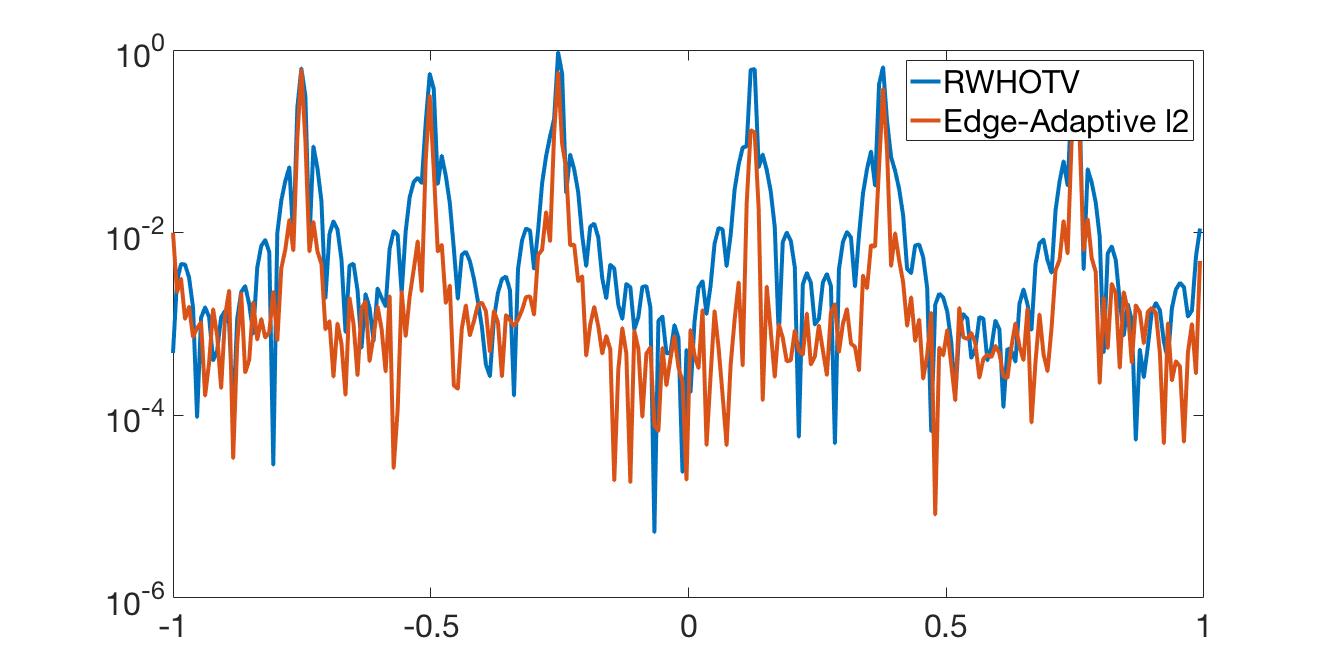}
\end{center}
\caption{Comparison of Algorithms \ref{alg:irl1} and \ref{alg:eal2} on $f_2(x)$ using (top) $m=1$ (middle) $m=2$ and (bottom) $m=3$ given $257$ jittered Fourier samples reconstructed on $257$ grid points. Left are the reconstructions, right are the corresponding pointwise errors. For parameters, we use $\rho=1$, $\ell_{max}=25$, $\epsilon=2.9$, $\mu=1$, $\tau=1/257$, and $\lambda=1$.}
\label{fig:1Dgelb}
\end{figure}
We now consider $f_2(x)$ in (\ref{eq:f2}). Because of the increased variation between the jumps, we test different values of $m$, the order of the PA method.  Recall that the polynomial annihilation (PA) method annihilates polynomials of degree $m - 1$ in smooth regions.  Figure \ref{fig:1Dgelb} compares the results using $m = 1,2,3$. We see improved overall accuracy and lower error around jumps using Algorithm \ref{alg:eal2} in each case.

\subsection*{One-dimensional examples with noise}
Noise in the data acquisition process of imaging systems has the potential to seriously degrade the quality of a reconstruction.  Figure \ref{fig:1Dnoise} compares each algorithm for both $f_1(x)$ and $f_2(x)$ when our data consists of  $257$ noisy jittered Fourier samples, (\ref{eq:noisydata}). Here the noise is assumed to be zero-mean complex Gaussian. While the edge-adaptive $\ell_2$ no longer provides significant improvement in the overall error, it is still evident that the functions are resolved better in the neighborhoods of the jumps.

\begin{figure}[h!]
\begin{center}
\includegraphics[width=.45\textwidth]{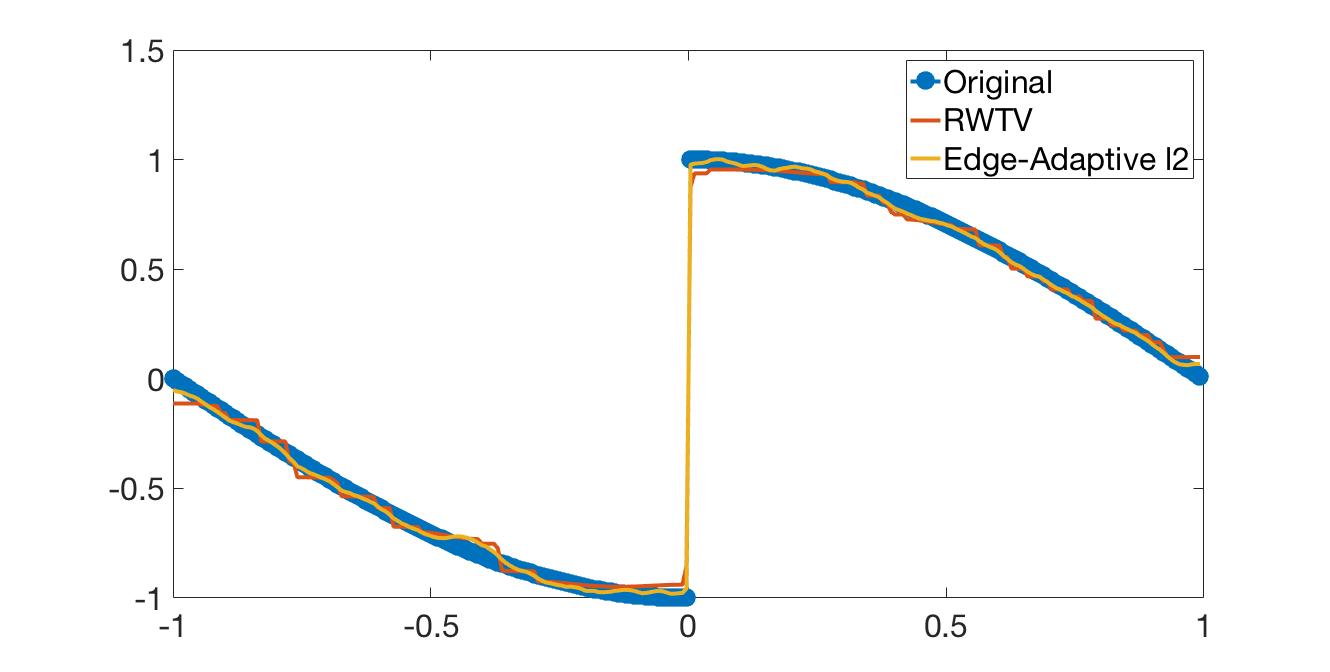}
\includegraphics[width=.45\textwidth]{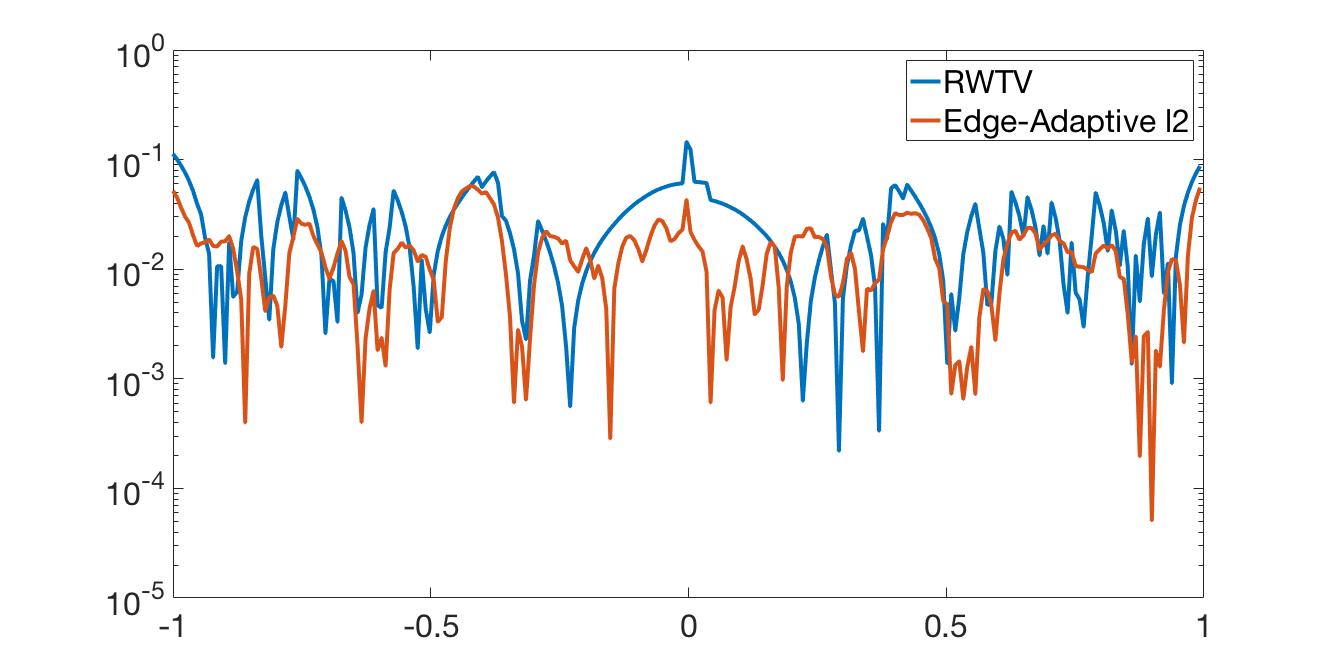}
\includegraphics[width=.45\textwidth]{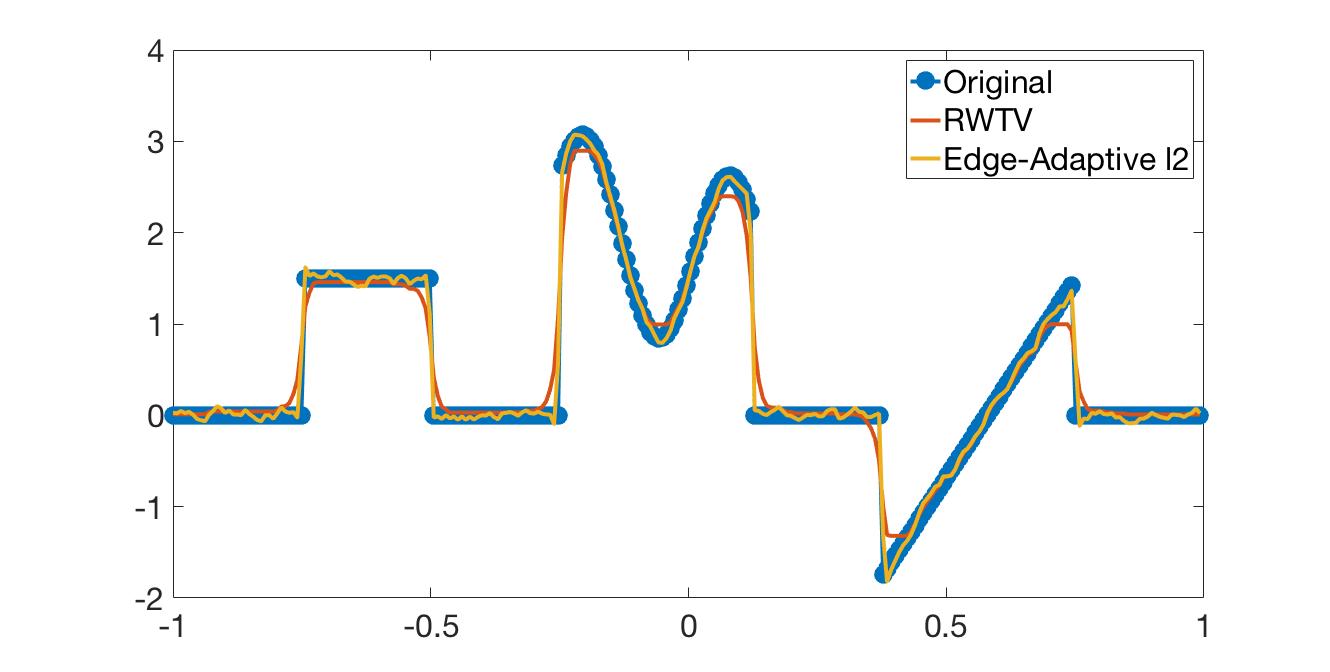}
\includegraphics[width=.45\textwidth]{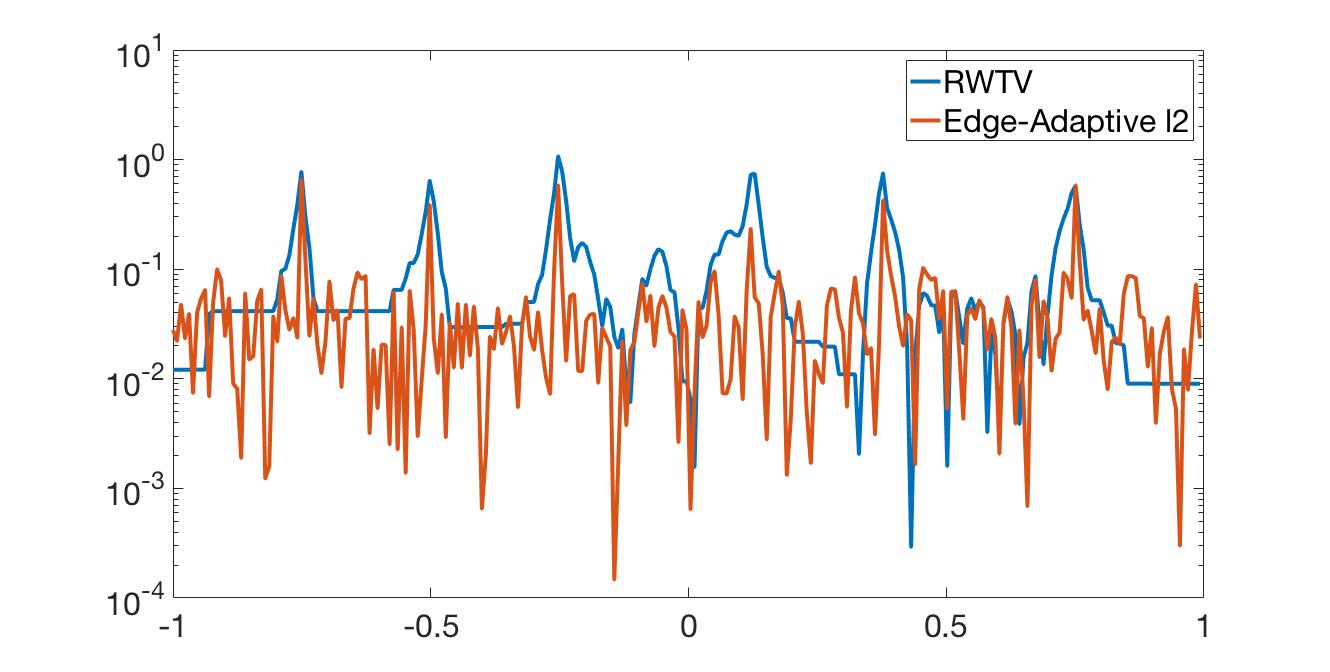}
\end{center}
\caption{(Left) Comparisons of Algorithms \ref{alg:irl1} and \ref{alg:eal2} on $f_1(x)$ and $f_2(x)$ given $257$ noisy jittered Fourier samples.   $SNR = 15$  for $f_1(x)$ and SNR $=20$ for $f_2(x)$. In both cases we use PA order $m = 1$ and reconstruct on $257$ grid points; (Right) corresponding pointwise errors. For $f_1(x)$ we use parameters $\rho=1$, $\epsilon=1.9$, $\ell_{max}=25$, $\mu=1$, $\tau=1/257$, and $\lambda=1$. For $f_2$ we use parameters $\rho=1$, $\epsilon=2.9$, $\ell_{max}=25$, $\mu=1$, $\tau=1/257$, and $\lambda=1$.}
\label{fig:1Dnoise}
\end{figure}

\subsection*{Limited data (compressed sensing) example}

\begin{figure}[htbp]
\begin{center}
\includegraphics[width=.35\textwidth]{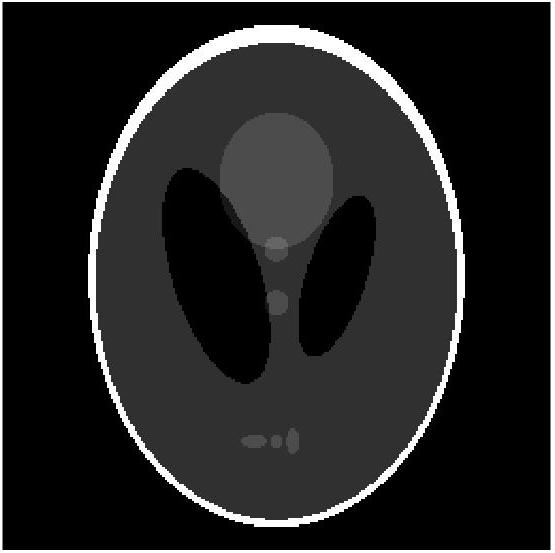} 
\end{center}
\caption{$257\times257$ pixel Shepp-Logan phantom.}
\label{fig:shepplogan}
\end{figure}

To test our algorithm's performance when starting from limited data, we consider the Shepp-Logan phantom, \cite{shepp1974fourier}, shown in Figure \ref{fig:shepplogan}. In this experiment, we randomly select just part of the initial Fourier data to use. Starting from $257\times257$ jittered Fourier modes as in (\ref{eq:fourierdata2d}), Figures \ref{fig:quarter}, \ref{fig:half} and \ref{fig:3quarters} respectively show the results using roughly a fourth, half, and three fourths of these modes. It is evident that the edge adaptive algorithm is particularly effective when the edges are close together, that is, it appears in general to have better resolution properties. Further theoretical and numerical study is needed to determine precisely the maximum compression ratio achievable by this method.\footnote{We note that typically Shepp Logan phantom reconstructions using compressive sensing algorithms come from either uniform or radial Fourier data, \cite{candes2006robust,candes2008enhancing}, which we are not considering here.}

\begin{figure}[htbp]
\begin{center}
\includegraphics[width=.30\textwidth]{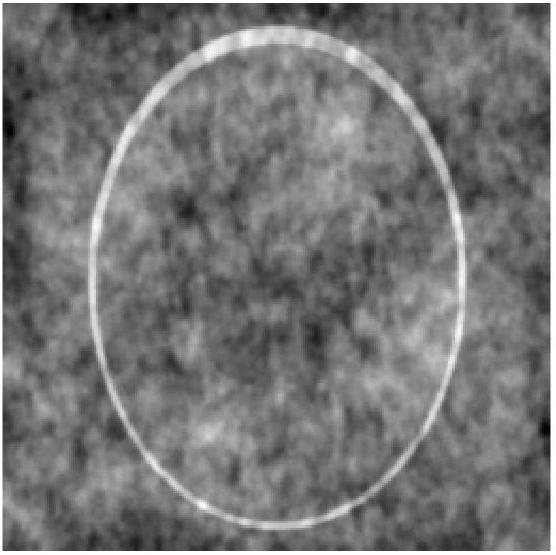}
\includegraphics[width=.30\textwidth]{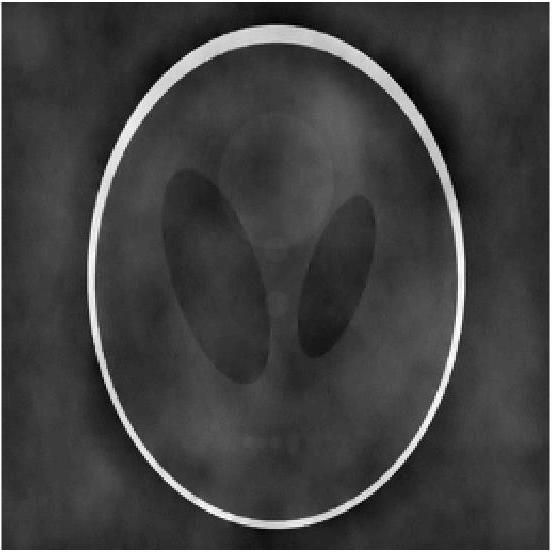}\\
\includegraphics[width=.31\textwidth]{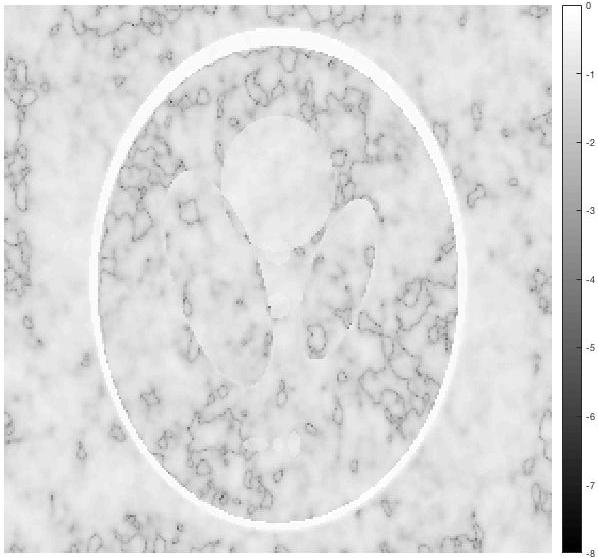}
\includegraphics[width=.31\textwidth]{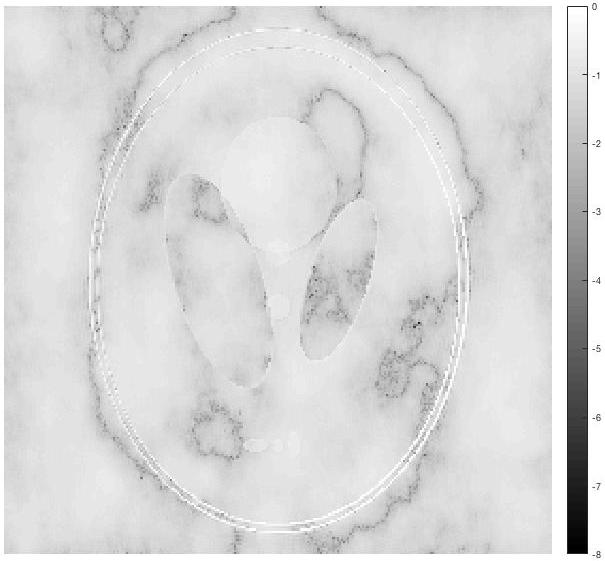}
\end{center}
\caption{(Top) Reconstruction on $257\times 257$ grid points of the Shepp-Logan phantom using Algorithms \ref{alg:irl12d} (left) and \ref{alg:eal22d} (right) with PA order $m=1$ from $129^2$ Fourier coefficients randomly chosen from a grid of $257\times257$. (Bottom) respective pointwise errors. For parameters, we use $\rho=.01$, $\epsilon=.9$, $\ell_{max}=5$, $\mu=.01$, $\tau=0.1$, $\lambda=.1$. The relative error using Algorithm \ref{alg:irl12d} was $RE=.7160$, while Algorithm \ref{alg:eal22d} yielded $RE=.4873$.}
\label{fig:quarter}
\end{figure}

\begin{figure}[htbp]
\begin{center}
\includegraphics[width=.30\textwidth]{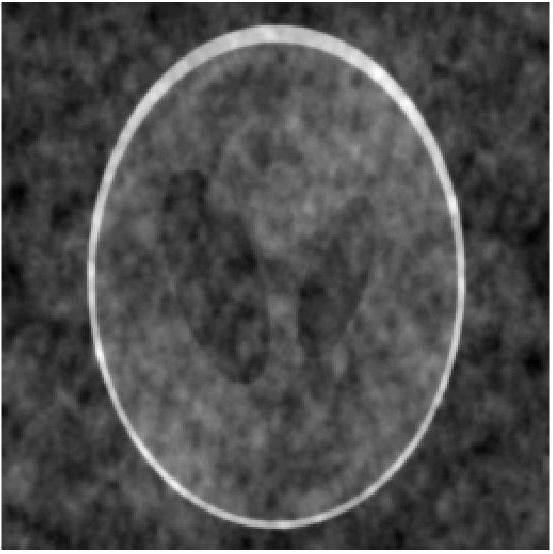}
\includegraphics[width=.30\textwidth]{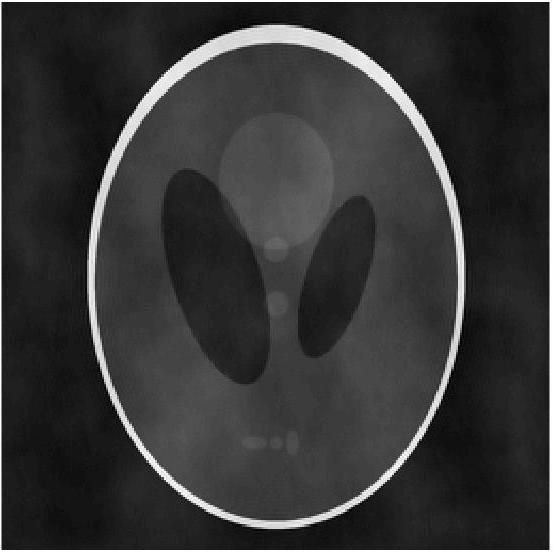}\\
\includegraphics[width=.31\textwidth]{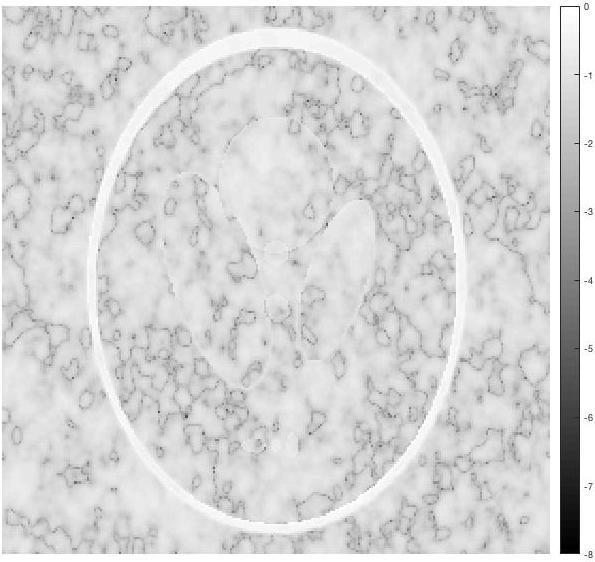}
\includegraphics[width=.31\textwidth]{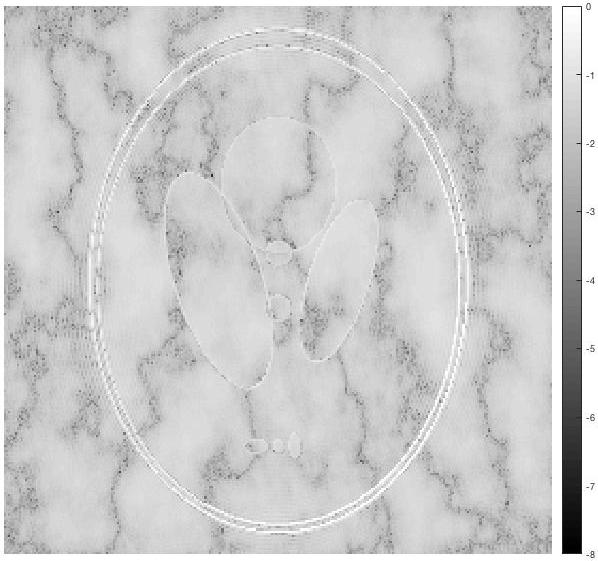}
\end{center}
\caption{(Top) Reconstruction on $257\times 257$ grid points of the Shepp-Logan phantom using Algorithms \ref{alg:irl12d} (left) and \ref{alg:eal22d} (right) with PA order $m=1$ from $181^2$ Fourier coefficients randomly chosen from a grid of $257\times257$. (Bottom) respective pointwise errors. For parameters, we use $\rho=.01$, $\epsilon=.9$, $\ell_{max}=5$, $\mu=.01$, $\tau=0.1$, $\lambda=.1$. The relative error using Algorithm \ref{alg:irl12d} was $RE=.5180$, while Algorithm \ref{alg:eal22d} yielded $RE=0.3259$.}
\label{fig:half}
\end{figure}

\begin{figure}[htbp]
\begin{center}
\includegraphics[width=.30\textwidth]{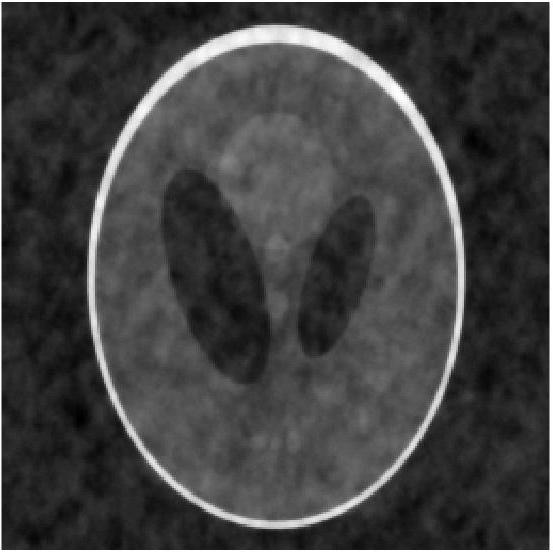}
\includegraphics[width=.30\textwidth]{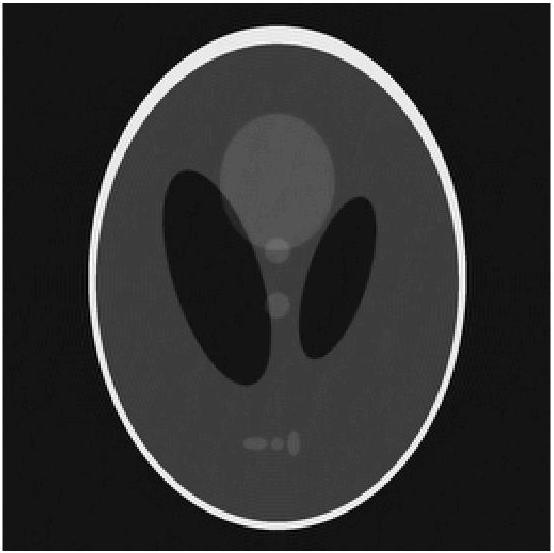}\\
\includegraphics[width=.31\textwidth]{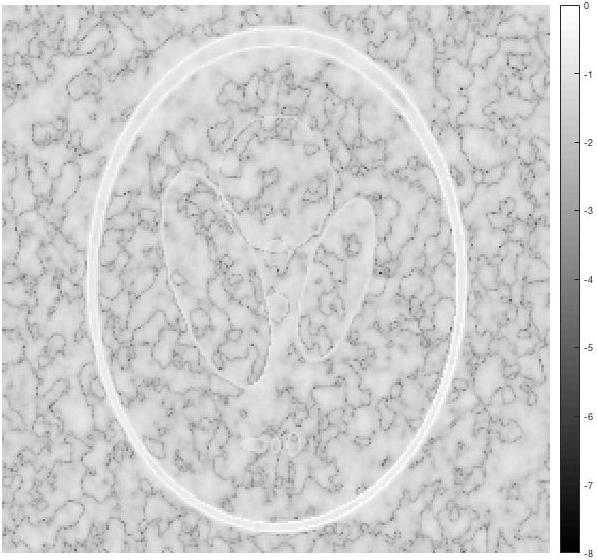}
\includegraphics[width=.31\textwidth]{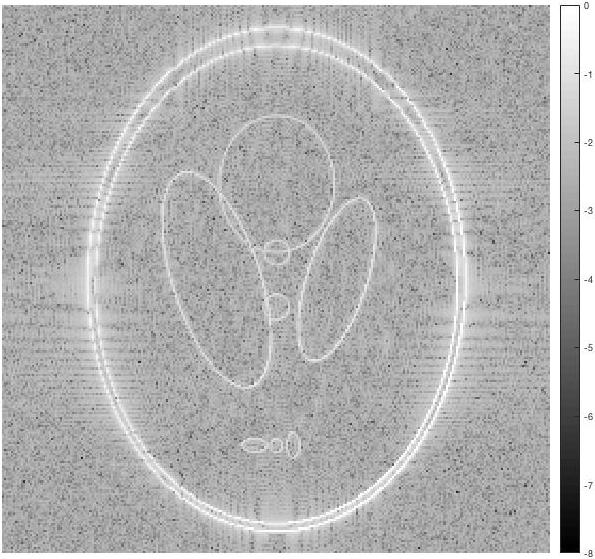}
\end{center}
\caption{(Top) Reconstruction on $257\times 257$ grid points of the Shepp-Logan phantom using Algorithms \ref{alg:irl12d} (left) and \ref{alg:eal22d} (right) with PA order $m=1$ from $225^2$ Fourier coefficients randomly chosen from a grid of $257\times257$. (Bottom) respective pointwise errors. For parameters, we use $\rho=.01$, $\epsilon=.9$, $\ell_{max}=5$, $\mu=.01$, $\tau=0.1$, $\lambda=.1$. The relative error using Algorithm \ref{alg:irl12d} was $RE=.3458$, while Algorithm \ref{alg:eal22d} yielded $RE=.2930$.}
\label{fig:3quarters}
\end{figure}

\subsection*{Synthetic Aperture Radar (SAR) example}

As a final example, we consider the synthetic aperture radar (SAR) phase history data of a vehicle given in \cite{dungan2010civilian}. SAR is an all weather, night or day imaging modality whereby an image is reconstructed from electromagnetic scattering data. In SAR we assume only a sparse number of isotropic point scatterers, so the standard $\ell_1$ regularized reconstruction solves
\begin{align}\label{eq:SARl1}
\mathbf{f}^* &= \arg\min_{\mathbf{g}}\left(||\mathcal{F}_{N}\mathbf{g}-\mathbf{\hat{f}}||_2^2+\lambda||\Theta^*\mathbf{g}||_1\right)
\end{align}
where $\Theta^*$ is a diagonal phase extraction matrix yielding $\Theta^*\mathbf{f}\approx|\mathbf{f}|$. This is needed because the phase of $\mathbf{f}$ is not sparse. (See e.g.~\cite{sanders2017composite} for details on the construction of $\Theta$.)  The edge-adaptive $\ell_2$ method for this application is then
\begin{align}
\mathbf{f}^* &= \arg\min_{\mathbf{g}}\left(||\mathcal{F}_{N}\mathbf{g}-\mathbf{\hat{f}}||_2^2+\lambda||M\mathbf{g}||_2^2\right)
\end{align}
where $M=M^x+M^y$ is the mask found through Algorithm \ref{alg:mask2d}. Since we are now using $\ell_2$ regularization, the phase extraction matrix $\Theta^*$ is unnecessary. SAR data have a significant amount of noise, \cite{franceschetti2018synthetic}. Nevertheless we are able to locate the edges with relatively high confidence.  Algorithm \ref{alg:eal22d} is particularly effective in this case because we can heavily penalize the regularization term, and for our experiments we chose $\lambda=100$. Figure \ref{fig:SAR} compares the results reconstructing via equation (\ref{eq:SARl1}) and Algorithm \ref{alg:eal22d} for the given SAR data set.

\begin{figure}[htbp]
\begin{center}
\includegraphics[width=.45\textwidth]{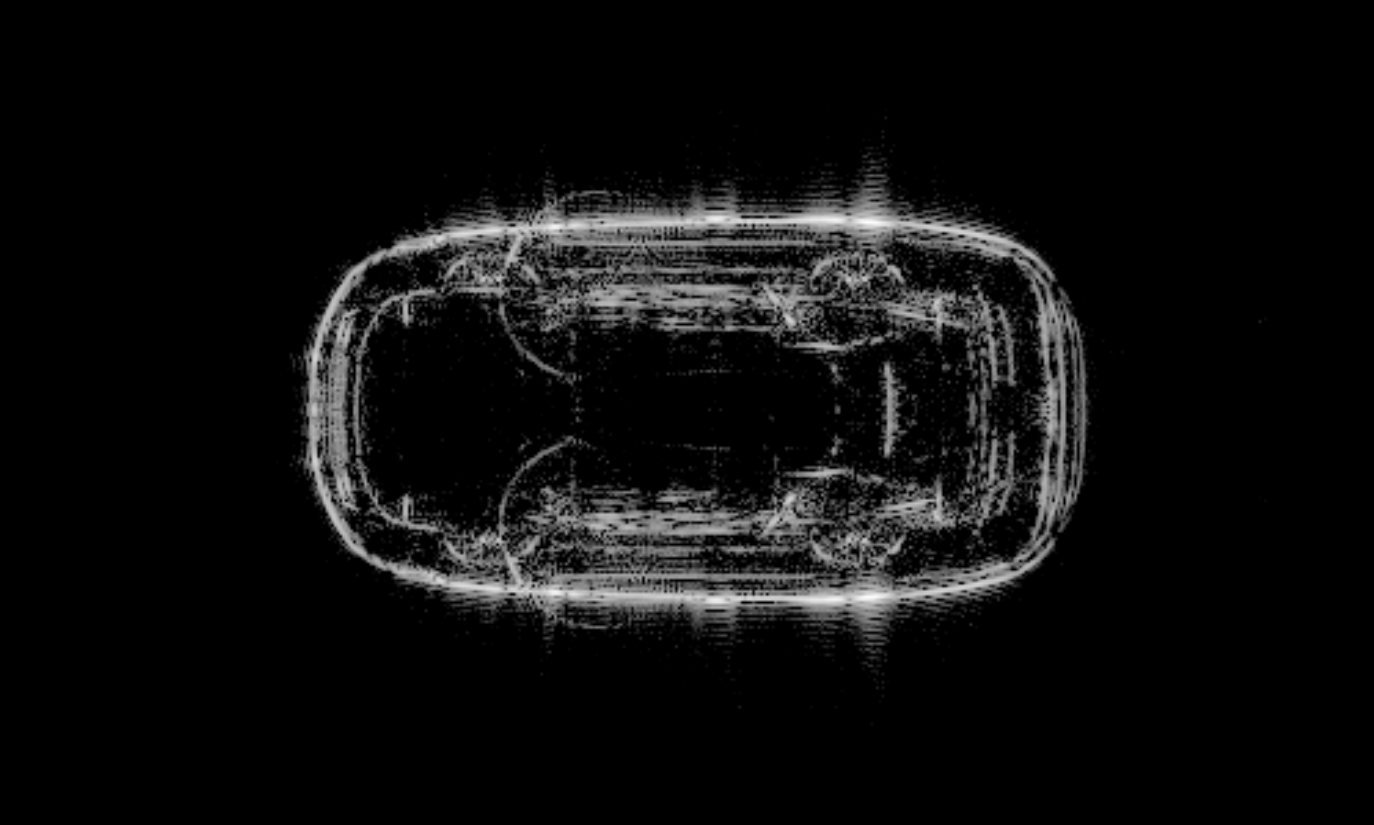} 
\includegraphics[width=.45\textwidth]{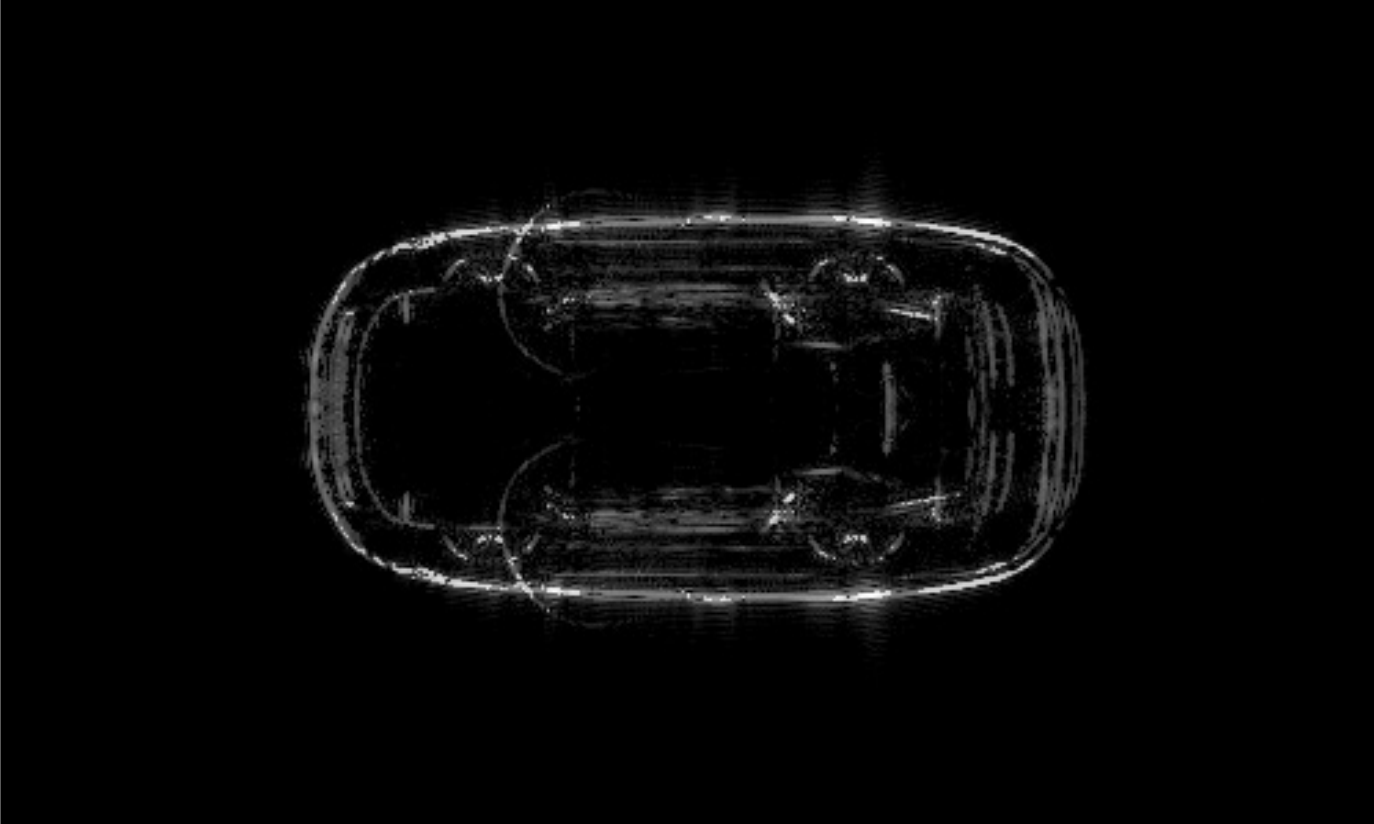} 
\includegraphics[width=.45\textwidth]{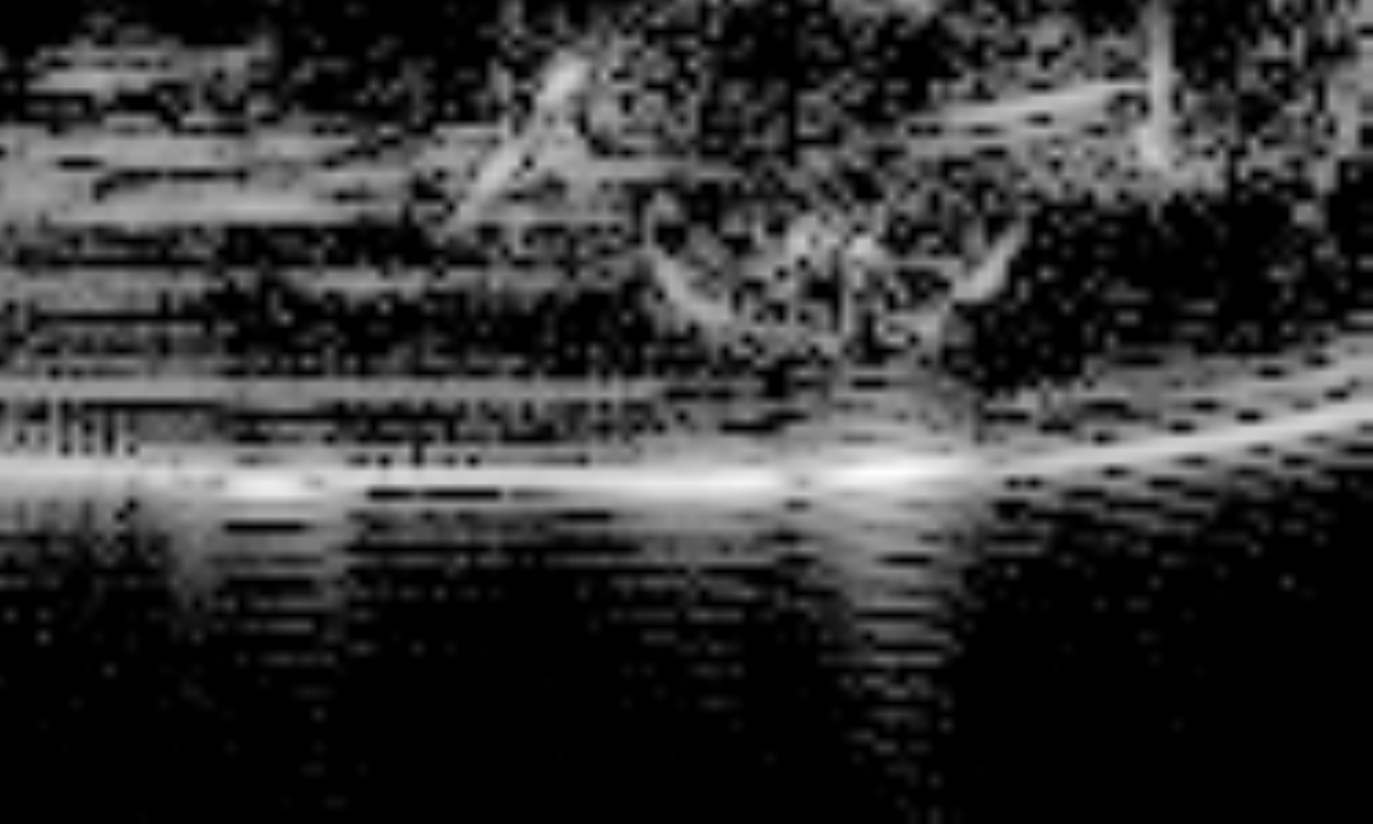} 
\includegraphics[width=.45\textwidth]{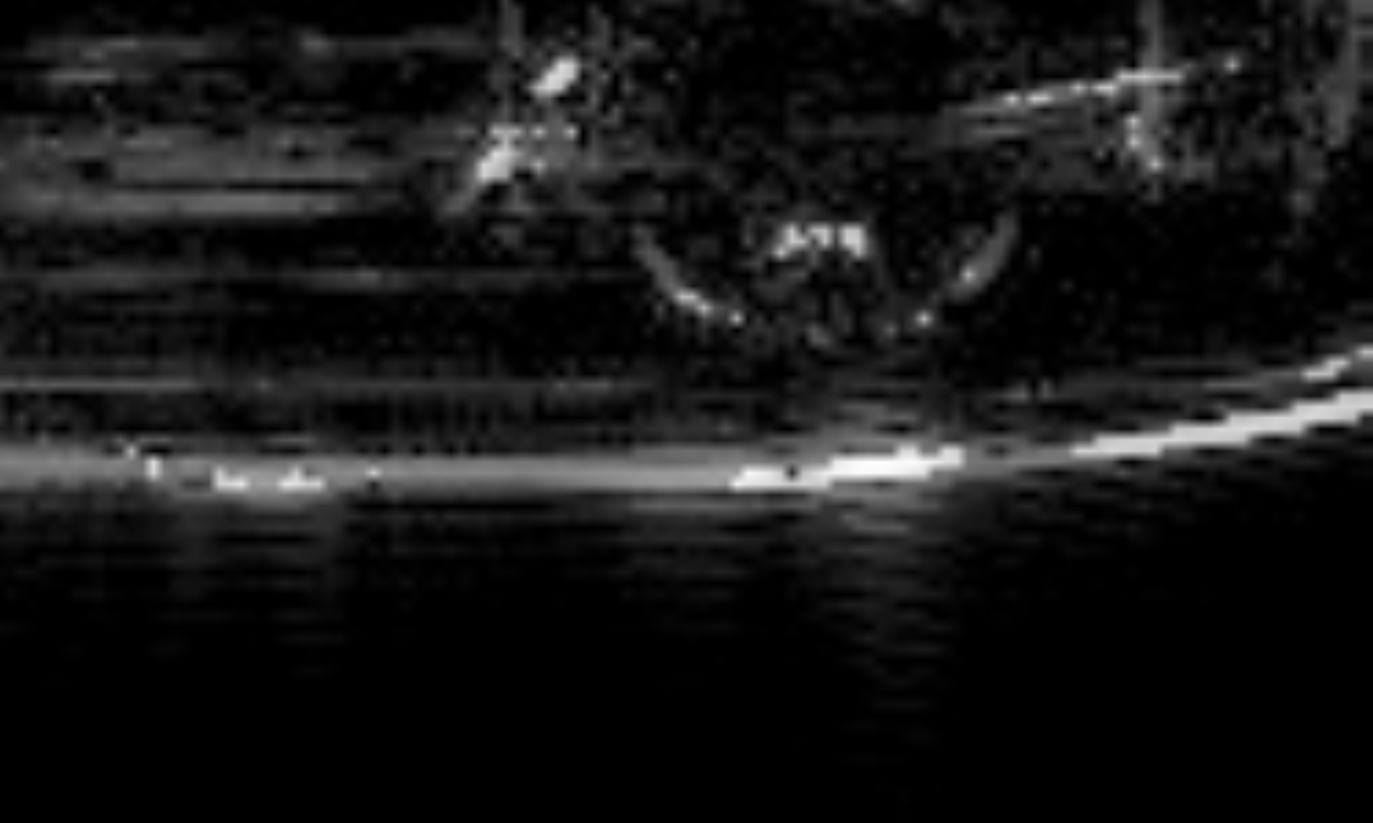} 
\end{center}
\caption{(Top) Reconstruction comparison of SAR vehicle data using (left) equation (\ref{eq:SARl1}) and (right) Algorithm \ref{alg:eal22d}. (Bottom) A close up of the lower right tire.}
\label{fig:SAR}
\end{figure}

\subsection*{Efficiency of Edge-Adaptive $\ell_2$ minimization}

Our new edge-adaptive $\ell_2$ method was shown to be more efficient than Algorithm \ref{alg:irl12d} in all experiments. This is to be expected since in general $\ell_2$ regularized problems are much easier to solve than $\ell_1$ minimizations.
 
\begin{table}[h!]
\centering
\begin{tabular}{|c |c| c|} 
 \hline
 Data size & Algorithm \ref{alg:irl12d} & Algorithm \ref{alg:eal22d} \\
 \hline
 $129\times129$ & 4 mins 10 secs & 5.6 secs  \\
 \hline
  $257\times257$ & 13 mins 2 secs & 22 secs  \\
 \hline
 $513\times513$ & 49 mins 26 secs & 1 min 33 secs  \\
 \hline
  $1025\times1025$ & 3 hours 16 mins & 6 mins 28 secs  \\
 \hline
\end{tabular}
\caption{Run time comparison between Algorithms \ref{alg:irl12d} and  \ref{alg:eal22d} for reconstructing $f_3(x,y)$.  We used $\ell_{max}=5$.  The run time includes the time to perform Algorithm \ref{alg:mask2d}.}
\label{table:1}
\end{table}
Table \ref{table:1} shows a comparison of the runtimes\footnote{All computations were performed on a MacBook Air with a 1.7 GHz Intel Core i5 processor and 4 GB of memory.} for Algorithm \ref{alg:irl12d} using $\ell_{max}=5$ and Algorithm \ref{alg:eal22d} including the mask generation in Algorithm \ref{alg:mask2d} for $f_3(x,y)$. For smaller images, e.g.~those reconstructed on a $129\times129$ pixel grid given $129\times129$ Fourier samples, the runtime for Algorithm \ref{alg:eal22d} is in seconds, compared to minutes for Algorithm \ref{alg:irl12d}. Note that this means that Algorithm \ref{alg:eal22d}, including edge detection, is faster than even a single iteration of Algorithm \ref{alg:irl12d}. These gains are even more significant as the images increase in size. For example, given $1025\times 1025$ Fourier samples reconstructed on $1025\times1025$ grid points, our new algorithm computes the results in about $6$ and a half minutes, while Algorithm \ref{alg:irl12d} took over $3$ hours, which is over $30$ minutes per iteration. We note that we did not implement accelerated homotopy-based algorithms for reweighted $\ell_1$ methods as in \cite{asif2013fast}, which may increase computational speed. In addition, the iteratively reweighted least squares method developed in  \cite{chartrand2008iteratively} would also run more efficiently, since it also uses an $\ell_2$ norm in the regularization. However, we would expect this method to also suffer from the same inaccuracies that arise from iteratively finding edges.

\section{Conclusion} \label{conclusion}
The edge-adaptive $\ell_2$ regularization image reconstruction method introduced in this paper compares favorably in terms of image quality, sharpness around jumps, and noise reduction to $\ell_1$ based and iteratively reweighted $\ell_1$ based regularization reconstructions. It is also more efficient, requiring just a single $\ell_1$ minimization solution that only needs to be performed once for an image. In fact, if the edges are already known from some other experiment, they can directly be used in our algorithm. After the edge detection, we can rely on faster conjugate gradient descent methods to solve the easier $\ell_2$ minimization problem. For some applications it may be useful to use the edge map produced in Algorithm \ref{alg:mask2d} as a cross-validation of the image.  The results for compressed imaging are promising, although more work is needed to determine how much compression is possible.  Future investigations will also include a variable (rather than binary) map, which may be important when the intensity of the images vary widely in scale. Moreover, this will allow us to generalize our technique to any problems for which separation of scales may be advantageous, that is, not just for identifying edges.  To this end, recent work in \cite{scarnati2018thesis} on weighted $\ell_p$ regularization methods might be useful.  We also will extend our new algorithm to multi-measurement vector (MMV) applications, as the efficiency gained by our method would be even more significant in this case.  Finally, we will test our method on other types of acquired data as well as other sparsifying transform operators, such as wavelets, which may be advantageous in some applications.

\bibliographystyle{acm}
\bibliography{refs}

\end{document}